\documentclass[letterpaper,english,reprint,aps,prl,superscriptaddress,showpacs,longbibliography]{revtex4-1}
\usepackage[latin9]{inputenc}
\usepackage{color}
\usepackage{amsmath}
\usepackage{amssymb}
\usepackage{graphicx}

\makeatletter

\pdfpageheight\paperheight
\pdfpagewidth\paperwidth

\DeclareTextSymbolDefault{\textquotedbl}{T1}
\newcommand{\lyxdot}{.}

\usepackage{babel}
\usepackage[colorlinks,
            linkcolor=red,
            anchorcolor=black,
            citecolor=blue
            ]{hyperref}

\makeatother

\usepackage{babel}
\begin{document}

\title{Experimental demonstration of switching entangled photons based on
the Rydberg blockade effect}

\author{Yi-Chen Yu}

\affiliation{Key Laboratory of Quantum Information, University of Science and
Technology of China, Hefei, Anhui 230026, China.}

\affiliation{Synergetic Innovation Center of Quantum Information and Quantum Physics,
University of Science and Technology of China, Hefei, Anhui 230026,
China.}

\author{Ming-Xin Dong}

\affiliation{Key Laboratory of Quantum Information, University of Science and
Technology of China, Hefei, Anhui 230026, China.}

\affiliation{Synergetic Innovation Center of Quantum Information and Quantum Physics,
University of Science and Technology of China, Hefei, Anhui 230026,
China.}

\author{Ying-Hao Ye}

\affiliation{Key Laboratory of Quantum Information, University of Science and
Technology of China, Hefei, Anhui 230026, China.}

\affiliation{Synergetic Innovation Center of Quantum Information and Quantum Physics,
University of Science and Technology of China, Hefei, Anhui 230026,
China.}

\author{Guang-Can Guo}

\affiliation{Key Laboratory of Quantum Information, University of Science and
Technology of China, Hefei, Anhui 230026, China.}

\affiliation{Synergetic Innovation Center of Quantum Information and Quantum Physics,
University of Science and Technology of China, Hefei, Anhui 230026,
China.}

\author{Dong-Sheng Ding}
\email{dds@ustc.edu.cn}

\affiliation{Key Laboratory of Quantum Information, University of Science and
Technology of China, Hefei, Anhui 230026, China.}

\affiliation{Synergetic Innovation Center of Quantum Information and Quantum Physics,
University of Science and Technology of China, Hefei, Anhui 230026,
China.}

\author{Bao-Sen Shi}
\email{drshi@ustc.edu.cn}

\affiliation{Key Laboratory of Quantum Information, University of Science and
Technology of China, Hefei, Anhui 230026, China.}

\affiliation{Synergetic Innovation Center of Quantum Information and Quantum Physics,
University of Science and Technology of China, Hefei, Anhui 230026,
China.}

\date{\today}
\begin{abstract}
\textcolor{black}{The}\textcolor{red}{{} }\textcolor{black}{long-range}
interaction between Rydberg-excited atoms endows a medium with large
optical nonlinearity. Here, we demonstrate an optical switch to operate
on a single photon from an entangled photon pair under a Rydberg electromagnetically
induced transparency configuration. With the presence of the Rydberg
blockade effect, we switch on a gate field to make the atomic medium
nontransparent thereby absorbing the single photon emitted from another
atomic ensemble via the spontaneous four-wave mixing process. In contrast
to the case without a gate field, more than 50\% of the photons sent
to the switch are blocked, and finally achieve an effective single-photon
switch. There are on average 1$\sim$2 gate photons per effective
blockade sphere in one gate pulse. This switching effect on a single
entangled photon depends on the principal quantum number and the photon
number of the gate field. Our experimental progress is significant
in the quantum information process especially in controlling the interaction
between Rydberg atoms and entangled photon pairs.
\end{abstract}
\maketitle
\begin{flushleft}
\par\end{flushleft}

\noindent \begin{flushleft}
\textbf{Pacs numbers: 32.80.Rm, 42.50.Gy, 42.50.Ct, 42.50.Nn}
\par\end{flushleft}

\noindent \begin{flushleft}
\textbf{1. Introduction}
\par\end{flushleft}

In analogy to classical electronic counterparts, quantum switches
are regarded as basic building blocks for quantum circuits and networks
\cite{cirac1997quantum,o2009photonic,caulfield2010future,Wang_2018}.
Switching states in the full-quantum regime where single particles
control a quantum qubit or entanglement from another system may enable
further applications in quantum information science, such as in quantum
computing \cite{saffman2010quantum}, distributed quantum information
processing \cite{kimble2008quantum,Cui_2019,Xie_2020}, and metrology
\cite{komar2014quantum}. Many efforts have been done towards constructing
a prototype; examples include a micro-resonator coupled with a single
atom \cite{o2013fiber}, cold atoms trapped in a microscopic hollow
fiber \cite{bajcsy2009efficient}, cold atoms coupled to a cavity
\cite{chen2013all}, strongly coupled quantum cavity-dots \cite{volz2012ultrafast},
and single dye molecules \cite{hwang2009single}.

The strong interaction offered by Rydberg-excited atoms shifts the
energy levels of the surrounding atoms dramatically and suppresses
all further excitation of these neighboring atoms. This interaction
between cold atoms gives rise to excitation blockade \cite{comparat2010dipole,jaksch2000fast,lukin2001dipole,tong2004local,singer2004suppression,urban2009observation,gaetan2009observation},
multipartite entanglement \cite{heidemann2007evidence,zeiher2015microscopic,labuhn2016tunable,bernien2017probing},
spatial correlations \cite{schauss2012observation,schwarzkopf2011imaging,schauss2015crystallization}\textcolor{black}{,
strong optical nonlinearities \cite{pritchard2010cooperative,dudin2012strongly,peyronel2012quantum,maxwell2013storage,tresp2015dipolar,firstenberg2016nonlinear,murray2017coherent},
plasma formation \cite{robert2013spontaneous}, and photon-photon
gate \cite{tiarks2019photon}. The single-photon nonlinearity arising
from the strong interaction between Rydberg atoms shows great potential
in constructing single-photon transistors \cite{tiarks2014single,gorniaczyk2014single,baur2014single}.
The fundamental aspects of quantum nonlinearity based on Rydberg atoms
have been studied before; a type of \textquotedbl photonic hourglass\textquotedbl{}
single-photon device is constructed \cite{firstenberg2013attractive,peyronel2012quantum}.
Such \textquotedbl photonic hourglass\textquotedbl{} is a kind of
medium exhibiting strong absorption of photon pairs while remaining
transparent to single photons. However, all the relative experiments
on switching were demonstrated with a weak coherent field, thus there
are no reports on switching of a true single-photon. Operating on
true single photons is more challenging than attenuated coherent pulses.}

\textcolor{black}{Here, we demonstrate an experiment of switching
true single-photons from entangle photon pairs. The entangled photon
pairs are prepared in one atomic cloud and propagate through another
atomic }cloud for switching operation. The switching operation is
using a single-photon level gate field to turn on or turn off the
acceptance of that true single-photon. We perpare a gate pulse with
1$\sim$2 photons per blockade sphere to switch one of the entangled
photons based on Rydberg blockade effect. By controlling the large
nonlinearity offered by the long-range interaction from the Rydberg
atoms, we can turn on and turn off the Rydberg electromagnetically
induced transparency (Rydberg-EIT) window corresponding to the switch
on and switch off operation\cite{pritchard2010cooperative,petrosyan2011electromagnetically}.
The measured coincidence counts with and without a gate field show
an obvious switching effect of the entangled photons. The fidelity
of the entangled state is $85.3\%\pm1.5\%$ and $80.4\%\pm2.3\%$
in the absence and presence of a gate field, respectively, with a
switch contrast larger than $50\%$. By increasing the principal quantum
number $n$, the switching effect becomes stronger and the required
photon number of the gate field decreases. Implementing a Rydberg-mediated
switch device under non-classical fields could enable the implementation
of quantum computation and information processing with the interaction
between Rydberg atoms and entangled photons \cite{saffman2010quantum},
such as building a Toffoli gate \cite{cory1998experimental} and quantum
computation \cite{gorshkov2011photon,khazali2015photon,wade2016single,sun2018analysis}
with Rydberg ensembles, and switching a distributed quantum node.
\begin{center}
\begin{figure*}[t]
\includegraphics[width=1\textwidth]{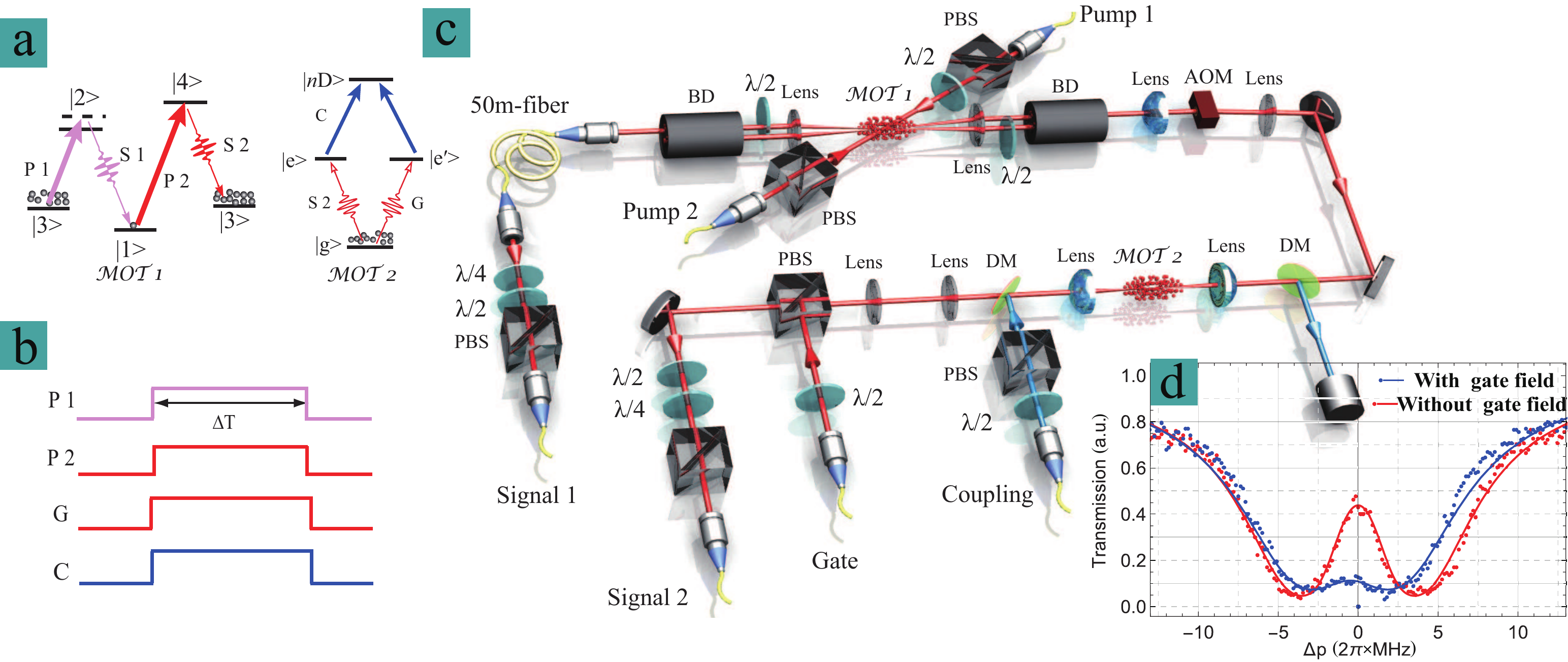}

\caption{(a) Energy diagram of entanglement generation and the switching processes.
The double-$\varLambda$ atomic configuration corresponds to the $^{85}Rb$
states $5S_{1/2}(F=2)$ ($\left|1\right\rangle $), $5P_{1/2}(F'=3)$
($\left|2\right\rangle $), $5S_{1/2}(F=3)$ ($\left|3\right\rangle $),
and $5P_{3/2}(F'=3)$ ($\left|4\right\rangle $), respectively. The
pump fields are P1 and P2 and the signal fields are S1 and S2. The
right-side energy diagram is a ladder-type energy level with ground
state $5S_{1/2}(F=3)$ ($\left|g\right\rangle $), excited state $5P_{3/2}(F'=4)$
($\left|e\right\rangle $), and highly-excited state $\left|nD_{5/2}\right\rangle $
($\left|nD\right\rangle $). Labels: P-pump, S-signal. (b) Time sequence
for the preparation and switching on single photons. $\Delta T$ represents
the experimental time window. (c) Schematic overview of the experimental
setup. Labels: PBS-polarizing beam splitter, DM-dichroic mirror, $\lambda/2$-half-wave
plate, $\lambda/4$-quarter-wave plate, BD\textendash beam displacer,
AOM\textendash acousto-optic modulator. (d) The Rydberg-EIT transmission
spectra is recorded with and without the gate field when $n=50$.
The solid lines are fitted by the function $e^{-2\mathrm{Im}[w_{s2}/c(1+\chi/2)]L}$
with $OD=8$, $\Omega_{c}=2\pi\times6.8$ MHz, $\delta\Delta=2\pi\times0$
MHz, $\gamma_{rg}=2\pi\times0.03$~MHz, $\text{\ensuremath{\mathbf{\Gamma}}}_{de}=2\pi\times0.07$
MHz and $OD=8$, $\Omega_{c}=2\pi\times5$ MHz, $\delta\Delta=2\pi\times0.5$
MHz, $\gamma_{rg}=2\pi\times2.5$~MHz for red and blue data, respectively.
The fitted function and these parameters are defined in theoretical
analysis part in \textcolor{black}{supplementary materials}.}

\label{experimental setup}
\end{figure*}
\par\end{center}

\noindent \begin{flushleft}
\textbf{2. Results}
\par\end{flushleft}

\noindent \begin{flushleft}
2.1. Experimental setup
\par\end{flushleft}

The sample media are optically thick atomic ensembles of Rubidium
85 ($^{85}Rb$) trapped in different magneto-optic traps (MOTs), labeled
MOT 1 and MOT 2. Schematics of the energy levels, time sequence, and
experimental setup are shown in figure~\ref{experimental setup}
(a)\textendash (c). A cigar-shaped $^{85}Rb$ atomic ensemble is first
prepared in MOT 1 and then cooled down to about 100~$\mu$K via the
optical molasses technique; the atomic cloud has dimensions of $10\times2\times2$~$\textrm{mm}^{3}$.
We prepare non-classical photon pairs by spontaneous four-wave mixing
(SFWM) in this atomic ensemble. The energy levels involved here correspond
to the double-$\Lambda$ system, consisting of both the D1 and D2
lines of Rubidium 85. The two pump fields couple the atomic transition
$5S_{1/2}(F=3)\rightarrow5P_{1/2}(F'=3)$ with a detuning of $-2\pi\times110$~MHz
and the atomic transition $5S_{1/2}(F=2)\rightarrow5P_{3/2}(F'=3)$
under resonance. The generated signal photons (labeled signal 1 and
signal 2) are correlated in the time domain. The signal-2 photon passing
through an acousto-optic modulator (AOM) is frequency shifted by $+2\pi\times120$~MHz,
after which it is exactly resonant with the atomic transition $5S_{1/2}(F=3)\rightarrow5P_{3/2}(F'=4)$.
Then, the signal-2 photon propagates through the three-dimensional
$^{85}Rb$ atomic cloud in MOT 2 for the demonstration of the switching
process. Finally, we perform quantum state tomography for the photonic
entanglement before two signals are detected by two single-photon
detectors. The coils of MOT 2 are switched off during the switching
measurement. The spherical atomic cloud of MOT 2 has a size of $500~\mu$m
with a temperature $\sim$ 20~$\mu$K and an average density of $3.5\times10^{11}\textrm{c\ensuremath{\textrm{m}^{-3}}}$.
The coupling field is resonant with the atomic transition $5P_{3/2}(F'=4)\rightarrow\left|nD_{5/2}\right\rangle $,
that is $\Delta_{c}=0$. Rabi frequency of coupling light is $\Omega_{c}=2\pi\times11$
MHz to demonstrate EIT and blockade configuration for single-photons.

The signal-2 photon has a beam waist of 16~$\mu\textrm{m}$ in the
center of MOT 2 which is obtained by using a short-focus lens. With
a pulsed coupling beam (TA-SHG, Toptica), we demonstrate Rydberg-EIT
in the ladder-type atomic configuration, consisting of a ground state
$\left|g\right\rangle $, an excited state $\left|e\right\rangle $,
and a highly-excited state $\left|nD\right\rangle $; here, $n=50$.
The gate field has a beam waist of 18~$\mu\textrm{m}$ in the center
of MOT 2 and couples the atomic transition $5S_{1/2}(F=3)\rightarrow5P_{3/2}(F'=4)$.
The coupling field with a beam waist of 30 $\mu\textrm{m}$ covers
both the gate and signal-2 beams. The smaller the beam waist, the
stronger the blockade effect \cite{peyronel2012quantum}. However
the beam size is limited to the size of the transmission window of
our MOT. Analyzing the van der Waals interactions between the Rydberg
atoms with an effective coefficient $\overline{C}_{6}=2\pi\times32$
$\textrm{GHz}\cdot\mu\mathrm{m}^{6}$ for the rubidium 50$D_{5/2}$
by considering weighted average of the interaction effects of all
Zeeman sublevels, we can calculate an average blockade radius $\sim$3.78~$\mu\textrm{m}$
with $\overline{r}_{b}=|\frac{\overline{C}_{6}}{\sqrt{(2\Delta_{c})^{2}+\Omega_{c}^{2}}}|^{1/6}$
\cite{balewski2014rydberg} \cite{vsibalic2017arc}. Since the coupling
Rabi frequency is larger than the bandwidth of the signal photon,
we use the strong coupling configuration to calculate the blockade
radius. Figure~\ref{experimental setup}(d) describes the switching
effect on a weak coherent pulse, the red and blue lines represent
the Rydberg-EIT spectra with and without a coherent gate field.

\begin{figure}[b]
\includegraphics[width=0.65\columnwidth]{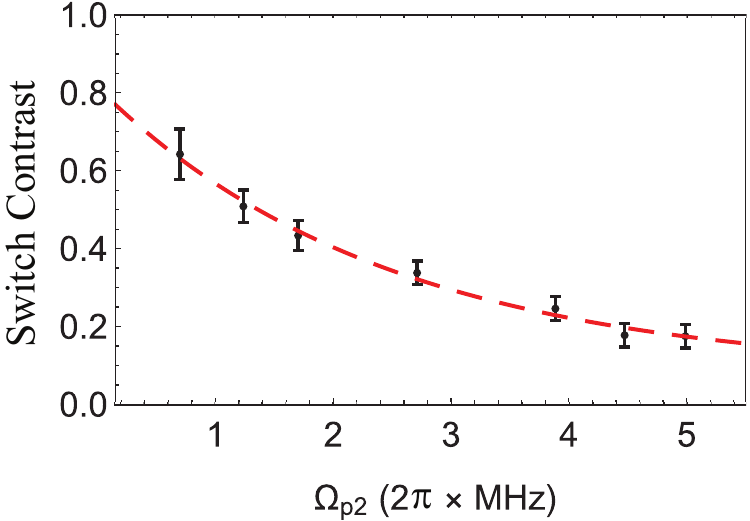}

\caption{The measured switch contrast against with $\Omega_{p2}$, the solid
red curve is guided for eye which is fitted by a function $y=A*\mathrm{Exp}[-x/t]+y_{0}$
with parameters of $A=0.73481;y_{0}=0.07865;t=2.45$.}

\label{contrast against pump2 }
\end{figure}

\noindent \begin{flushleft}
2.2. Bandwidth matching
\par\end{flushleft}

In order to switch single photons, we need to connect two physical
systems. One is to generate entangled photon; the other is to operate
on that photon. We match the frequency and the bandwidth between the
signal-2 photon and the absorption window of the atomic ensemble in
MOT 2. This can be realized by changing the frequency and Rabi frequency
$\Omega_{p2}$ of the pump 2 field as explained above. The switching
effect obviously decreases when the bandwidth of signal-2 photon increases.
Due to the narrow transparency window in the spectrum of Rydberg-EIT,
the optical response on two-photon resonance is strongly affected
by the level shifts induced by Rydberg atoms interaction and the linewidths
of the input lasers \cite{levine2018high}. Compared with $S$ state,
$D$ state has wider transparency window resulting from the larger
dipole matrix element to $D$ state. Thus, we use Rydberg-$D$ state
to get larger bandwidth and higher transmission rate of Rydberg-EIT
window.

As a result, the mismatching between the signal-2 photon and the absorbtion
bandwidth of the atomic ensemble in MOT 2 decreases the switch contrast.
Because the high-frequency component of the signal-2 photon is unable
to fall within the Rydberg-EIT window, the reabsorption of the signal-2
photon weakens although the gate field is present. The switch contrast
decreases with increasing Rabi frequency $\Omega_{p2}$ of pump 2
field (see figure~\ref{contrast against pump2 }). The bandwidth
of the signal-2 photon depends significantly on $\Omega_{p2}$ \cite{du2008narrowband,liao2014subnatural},
because the profile of the wave packet of the signal-2 photon can
be modulated by tuning the $\Lambda$-EIT transparency window. The
single-photon bandwidth becomes narrower with the $\Omega_{p2}$ decreasing.
Only when the bandwidth of the single photon is narrower than the
Rydberg-EIT window, can we get a higher absorption rate and switch
contrast.This data in figure~\ref{contrast against pump2 } hints
that the switching effect becomes more obvious with a smaller $\Omega_{p2}$.
For the optimized case $\Omega_{p2}\sim2\pi\times1$ MHz, the bandwidth
of the signal-2 photon is at $\sim2\pi\times5$~MHz, and the absorption
window for the atomic ensemble in MOT 2 is $\sim2\pi\times13$~MHz.
That is, the signal-2 photon can completely fall within the Rydberg-EIT
window. If decreasing $\Omega_{p2}$ further, the signal to noise
ratio of two-photon coincidence becomes worse.

\begin{figure}[b]
\includegraphics[width=1\columnwidth]{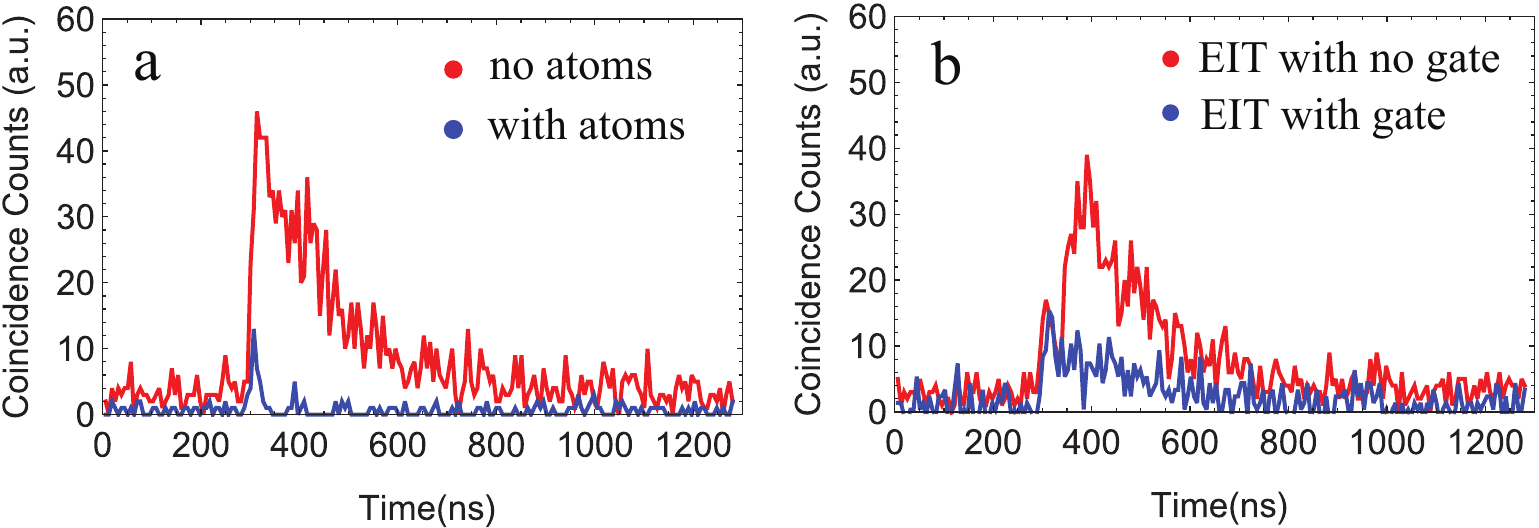}

\caption{Coincidence counts (CC) under different situations corresponding to:
(a) with (blue) and without atoms (red), and (b) under the Rydberg-EIT
configuration with (blue) and without (red) gate field. All of these
CC were detected in 4000~s.}

\label{blockade effect}
\end{figure}

\noindent \begin{flushleft}
2.3. Switch entangled photons
\par\end{flushleft}

To demonstrate the switching effect under quantum regime, we firstly
prepared non-classical photon pairs via SFWM process \cite{du2008narrowband,liao2014subnatural}
in MOT 1. The generated photon pairs are correlated in time domain.
We construct two optical paths $L$ and $R$ by using two beam displacers
(BDs) to build a passive-locking interferometer \cite{ding2016entanglement,zhang2016experimental,yu2018self}
where the perturbations between $L$ and $R$ optical paths can be
mutually eliminated. The signal photons in each path are collinear,
as the phase matching condition $k_{\mathrm{p1}}-k_{\mathrm{s1}}=k_{\mathrm{p2}}-k_{\mathrm{s2}}$
should be satisfied in the SFWM process. With two half-wave plates
inserted in the $R$ optical path, the signal photons along these
two optical paths can be coherently combined by BDs. The form of the
entanglement is
\begin{equation}
\left|\psi\right\rangle =(\left|H_{s1}\right\rangle \left|V_{s2}\right\rangle +e^{i\theta}\left|V_{s1}\right\rangle \left|H_{s2}\right\rangle )/\sqrt{2}
\end{equation}
with \textit{$\theta$}, the relative phase between $L$ and $R$
optical paths, setting to zero in our experiment; $\left|H_{s1,s2}\right\rangle $
and $\left|V_{s1,s2}\right\rangle $ represent the horizontal and
vertical polarized states of the signal photons. The details about
generating entangled photon pair are in our supplementary materials.

In order to demonstrate the switching effect of entangled photons,
we input entangled photons into MOT 2. In this situation, we use a
50-m fibre to introduce a time delay in the path of the signal-1 photons.
This guarantees that the entanglement does not collapse before the
switching process has finished. The results are shown in figure~\ref{blockade effect}(a)
and (b); the former shows the coincidence counts of photon pairs when
the atoms in MOT 2 are absent (red) and present (blue), whereas the
latter shows the results under Rydberg-EIT without (red) and with
(blue) gate field. Obviously, the coincidence counts decrease when
the gate field is applied as the signal-2 photon is significantly
absorbed when compared with the no-gate situation. The central physics
behind the operation of a single-photon switch is that the long-range
interaction between Rydberg atoms endows the Rydberg-EIT medium with
a large optical nonlinearity \cite{Pritchard2010,peyronel2012quantum},
and the resulting dipole blockade effect makes the medium non-transparent.
We define a switch contrast to characterize the switching effect,
\begin{equation}
\mathrm{C_{switch}}=\frac{\mathrm{CC_{EIT}}-\mathrm{CC_{gate}}}{\mathrm{CC_{EIT}}},
\end{equation}
where $\mathrm{CC_{EIT}}$ and $\mathrm{CC_{gate}}$ represent the
total coincidence counts between the signal-1 and signal-2 photons
without and with a gate field. From the data {[}figure~\ref{blockade effect}
(a) and (b){]}, we obtain a switch contrast of $\mathrm{C_{switch}}=77.6\%\pm3.1\%$.
The little peak in the rising edge comes from the high-frequency components
of the single photons, which falls out of the absorption window of
the atoms, {[}marked in blue color in figure~\ref{blockade effect}
(a){]}. In our experiment, the absorption window of atoms in MOT 2
is about $\sim2\pi\times13$~MHz. Although the bandwidth of the signal-2
photon wave-packet may be tuned by decreasing the power of pump 2
field \cite{liao2014subnatural}, there is always a high-frequency
component in the wave-packet of the signal-2 photon. And the component
falls outside of the bandwidth of the absorption, which induces an
optical precursor \cite{zhang2011optical,ding2015optical}. The switch
contrast is also limited by the broadening effect of the Rydberg-EIT
window, which is maybe caused by the dephasing of the distribution
of Rabi frequencies with the unpolarized atoms in MOT 2. Our experiment
is demonstrated with no bias magnetic fields. The atoms can be treated
as averagely distributed in all sublevels.
\noindent \begin{flushleft}
\begin{figure*}[t]
\includegraphics[width=1.8\columnwidth]{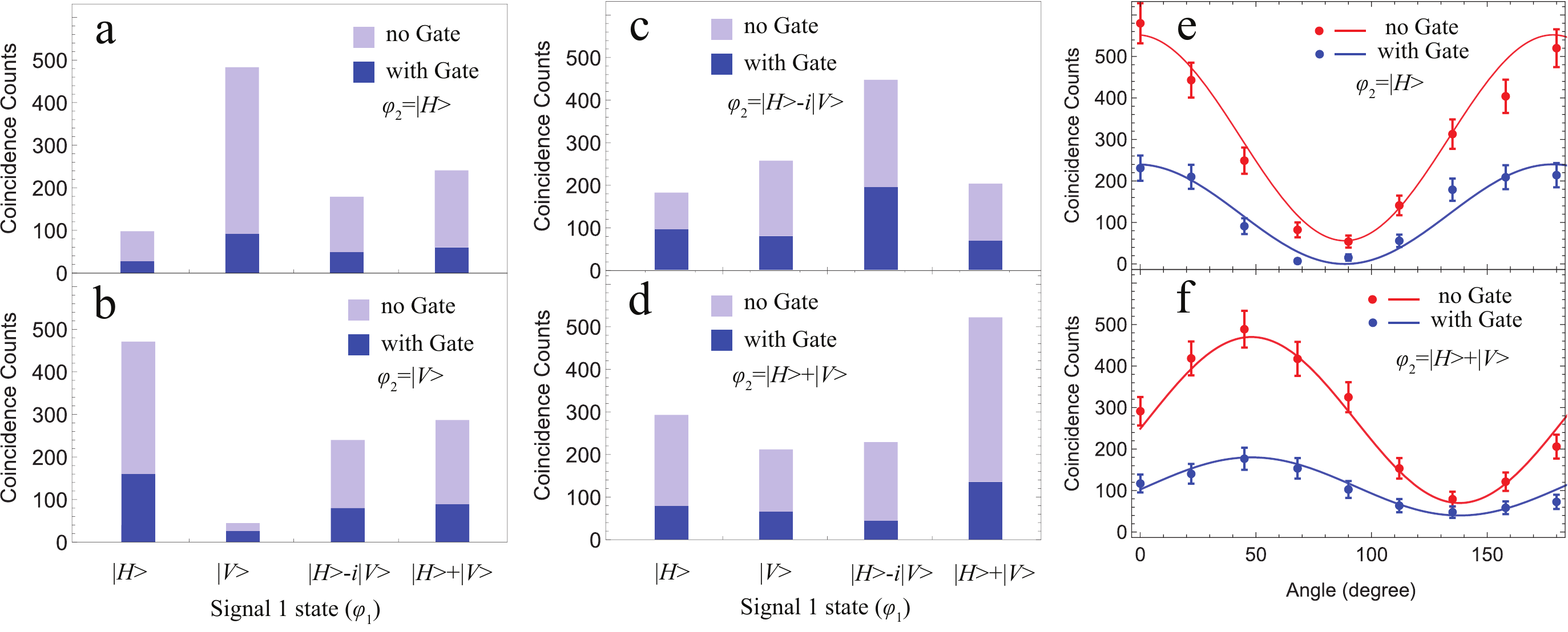}

\caption{Switching with different bases: (a)\textendash (d) are the coincidence
counts without a gate field (semi-transparent blue column) and with
a gate field (blue column) under different signal-1 states of $\left|H\right\rangle $,
$\left|V\right\rangle $, $\left|H\right\rangle -i\left|V\right\rangle $,
and $\left|H\right\rangle +\left|V\right\rangle $. (e) and (f) are
the recorded two-photon interference curves with signal-2 states of
$\left|H\right\rangle $ and $\left|H\right\rangle +\left|V\right\rangle $
without gate field (red) and with gate field (blue). Their interference
visibilities are $87.0\%\pm0.8\%$ ((e), red), $82.9\%\pm0.7\%$ ((e),
blue), $72.0\%\pm1.1\%$ ((f), red), and $57.1\%\pm2.6\%$ ((f), blue).
The solid lines are fitted curves to the measured data. All these
coincidence counts were recorded over a interval of 1000~s. Error
bars are $\pm1$ standard deviation.}

\label{counts and inteference}
\end{figure*}
\par\end{flushleft}

We change the detected state $\varphi_{\mathrm{s2}}$ of the signal-2
photon and recorded the coincidence counts under different signal-1
states $\varphi_{\mathrm{s1}}$ of $\left|H\right\rangle $, $\left|V\right\rangle $,
$\left|H\right\rangle -i\left|V\right\rangle $, and $\left|H\right\rangle +\left|V\right\rangle $.
To obtain the differences with and without the gate field, we recorded
these coincidence counts under these situations {[}figure~\ref{counts and inteference}
(a)\textendash (d){]}. The coincidence counts without (semi-transparent
blue)/with (blue) the gate field are obviously different. We obtain
switch contrasts with $\mathrm{C}_{\mathrm{switch}}^{VH}=81.0\%$,
$C_{\mathrm{switch}}^{HV}=64.3\%$, $C_{\mathrm{switch}}^{RR}=52.2\%$,
and $C_{\mathrm{switch}}^{DD}=79.9\%$ under the four situations $\varphi_{\mathrm{s1}}=\left|V\right\rangle $,
$\varphi_{\mathrm{s2}}=\left|H\right\rangle $; $\varphi_{\mathrm{s1}}=\left|V\right\rangle $,
$\varphi_{\mathrm{s2}}=\left|H\right\rangle $; $\varphi_{\mathrm{s1}}=\varphi_{\mathrm{s2}}=\left|H\right\rangle -i\left|V\right\rangle $
and $\varphi_{\mathrm{s1}}=\varphi_{\mathrm{s2}}=\left|H\right\rangle +\left|V\right\rangle $.
The switching operation is effective for any polarization state with
treating the signal-2 photon as if it were in a mixed state. The obtained
switching contrasts are different depending on the detected states
due to the non-perfect balance of the photon generation rate in the
two optical paths and the noise of each path. In addition, we measure
two-photon interference without and with the gate field und\textcolor{black}{er
the signal-2 basis of $\left|H\right\rangle $ and $\left|H\right\rangle +\left|V\right\rangle $
{[}figure~\ref{counts and inteference} (e) and (f){]}. From figure~\ref{counts and inteference}
(e), we find the visibility without the gate field exceeding the threshold
70.7\%, which means that the entanglement can be preserved after the
transmission through the EIT window. It is easy to observe that the
propagation of the signal-2 photon through the Rydberg-EIT medium
doesn't destroy the entanglement, because the Rydberg-EIT is independent
on the polarization of the signal-2 photon. }

\textcolor{black}{We also perform quantum state tomography \cite{james64measurement}
for the photonic entanglement to compare the entanglement properties
before and after the switching process. Signal-1 and signal-2 are
polarization entangled, their entangled state being $\left|\psi\right\rangle {\rm =}\left(\left|H\right\rangle _{\mathrm{s1}}\left|V\right\rangle _{\mathrm{s2}}+\left|V\right\rangle _{\mathrm{s1}}\left|H\right\rangle _{\mathrm{s2}}\right)/\sqrt{2}$.
Using the polarizing beam splitter, half-wave plate, and quarter-wave
plate, we project the two photon states onto the four polarization
states $\left|\phi_{1\sim4}\right\rangle $ ($\left|H\right\rangle $,
$\left|V\right\rangle $, $\left(\left|H\right\rangle -i\left|V\right\rangle \right)/\sqrt{2}$,
}$\left(\left|H\right\rangle +\left|V\right\rangle \right)/\sqrt{2}$).
We obtain a set of 16 data points from which to reconstruct the density
matrix. By comparing with the ideal density matrix $\rho_{\mathrm{ideal}}$
using the formula $F_{i}{\rm {=Tr(}}\sqrt{\sqrt{\rho}{\rho_{{\rm {ideal}}}}\sqrt{\rho}}{{\rm {)}}^{2}}$,
we obtain the fidelity to be $87.1\%\pm2.5\%$ corresponding to the
fidelity of input photons. By comparing with the input density matrix
using the formula of $F{\rm {=Tr(}}\sqrt{\sqrt{\rho_{\mathrm{output}}}{\rho_{{\rm {input}}}}\sqrt{\rho_{\mathrm{output}}}}{{\rm {)}}^{2}}$,
the fidelity for the output state without gate field is $85.3\%\pm1.5\%$,
and the state with gate field is $80.4\%\pm2.3\%$. The switch contrast
does not dramatically decrease before and after switch process. We
use the initial fidelity for the input state to characterize the entanglement
of the entangled photon pairs from SFWM process. Then, we want to
demonstrate that our switching operation is a quantum process without
any destructive effect to the input entanglement by calculating the
fidelity for the output state with and without the gate field, respectively.
We can still get high fidelity of the entanglement via much longer
detecting time after the transmission through the EIT window, as if
we hadn\textquoteright t operated on it.

The nonlinearity of the medium not only depends on the atomic density,
which determines the interaction distance, but also is strongly affected
by the dipole interaction strength. We change the principal quantum
number $n$ to change the interaction strength to measure both the
Rydberg-EIT transmission contrast and the switch contrast. Here the
Rydberg-EIT contrast is defined as $\mathrm{C_{EIT}=(CC_{no_{-}atom}-CC_{EIT})}\mathrm{/CC_{no_{-}atom}}$,
$\mathrm{CC_{no_{-}atom}}$ representing the total coincidence counts
between the signal-1 and signal-2 photons without atoms. We change
the principal quantum number $n$ by changing the wavelength of the
coupling laser. Each time we change the wavelength, we adjust the
experimental optical system to keep the coupling Rabi frequency a
constant. The results (figure~\ref{Against n}) show that the Rydberg-EIT
contrast decreases with the increase of $n$; this is because the
transition amplitude for $\left|e\right\rangle \rightarrow\left|nD\right\rangle $
decreases. In contrast, because the dipole interaction strength increases,
the switch contrast of the signal-2 photon obviously increases by
comparing two situations, $n=60$ ($\overline{r}_{b}=5.12$~$\mu\textrm{m}$)
and $n=40$ ($\overline{r}_{b}=1.05$~$\mu\textrm{m}$). The switch
contrast is larger than 50\% when $n>45$, revealing an effective
switching operation. In this way, the interaction between Rydberg
atoms becomes stronger with the principal quantum number increasing.
Although the gate field has hundreds of photons because of the relatively
large size of the atomic cloud in our experiment, we calculate that
there are on average $1\sim2$ gate photons per effective blockade
sphere corresponding to the average energy of the gate field inside
a blockade sphere expressed in units of $\hbar*\omega$. Eventually,
the single-photon switch with a single gate photon can be realized
by trapping the atoms into the scale of the blockade radius and increasing
the principal quantum number $n$ .
\noindent \begin{flushleft}
\begin{figure}[t]
\includegraphics[width=0.8\columnwidth]{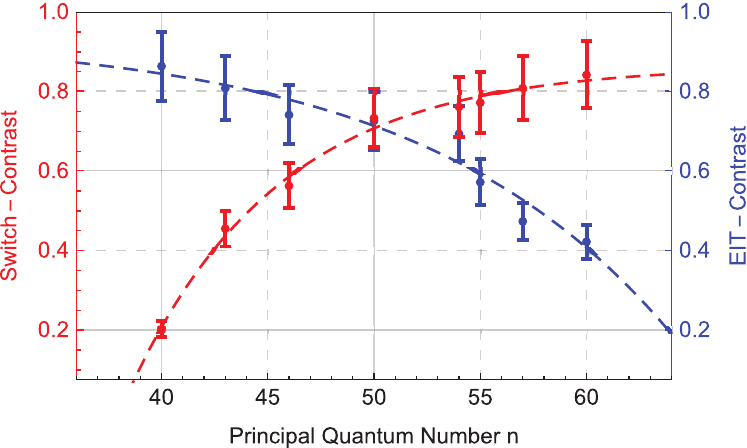}

\caption{Dependence of the switch contrast (red) and Rydberg-EIT contrast (blue)
on the principal quantum number. As visual guides, the data are fitted
with function $y=A*\mathrm{Exp}[-x/t]+y_{0}$ (dotted lines) with
parameter settings $A=-0.00327$, $y_{0}=0.94$, $t=-11.77$ and $A=-198.23$,
$y_{0}=0.87$, $t=7.0$. In this process, the gate field intensity
is set to average\textcolor{red}{{} }\textcolor{black}{$\sim$ 2} gate
photons per effective blockade sphere. Error bars are $\pm1$ standard
deviation.}

\label{Against n}
\end{figure}
\textbf{3. Discussion}
\par\end{flushleft}

In addition, in order to avoid saturation of our switch system where
the gate and coupling field would deplete the atoms after a certain
duration, we set the experimental time window to 25~$\mu$s. The
lifetime of Rydberg-excited atom is estimated to be $\sim$ 800 ns
by measuring the Rydberg spinwave through storage process \cite{ding2016entanglement,wang2017optical},
which guarantees an adequate interaction time for each switch operation
during near 200 ns arrival time of the signal-2 photon given in figure
\ref{blockade effect}. In our experiment, the dephasing of Rydberg
$D$ state \cite{tresp2015dipolar} is not obvious due to the small
gate photon numbers used here. With small photon numbers as input,
we have not investigated an obvious time dependence of the transmission
on Rydberg-EIT resonance, but with an obvious time dependence of the
transmission for large photon numbers, more details are shown in \textcolor{black}{supplementary
materials} figure 2. Thus, the nonlinearity behind the switch experiment
is offered by Rydberg blockade effect (see more details in \textcolor{black}{supplementary
materials} figure 3). Besides, there are some challenges to improve
the switch contrast: 1. Decreasing the bandwidth mismatch between
signal-photon and the Rydberg-EIT transparency window. 2. Increasing
the principle quantum number $n$ to achieve large dipole-dipole interaction
strength, as shown in figure \ref{Against n}. 3. Trapping the atoms
into the scale of the blockade radius. 4. Using a single-photon with
ultra-narrow bandwidth. The bandwidth of transparency window of Rydberg-EIT
with large principle quantum number $n$ would become narrower due
to smaller natural line width for high-$n$ Rydberg state. This means
that it needs narrower bandwidth of single-photon, such as subnatural-linewidth
single-photon source \cite{liao2014subnatural}.

In summary, we have demonstrated an optical switch on entangled photons
based on Rydberg nonlinearity with two atomic ensembles. The emitted
signal-2 photon correlated with the signal-1 photon is blocked by
another gate field under the Rydberg-EIT configuration. Switching
effect depends on the principal quantum number, the bandwidth of the
emitted single photons, and the average photon number of the gate
field. We have successfully realized optical switch on one of the
entangled photons of the pair, with more than 50\% of pairs being
blocked. These results on switching single photons using the strong
dipole interaction hold promises in demonstrating quantum information
processing between Rydberg atoms and entangled photons.
\noindent \begin{flushleft}
\textbf{4. Acknowledgements}
\par\end{flushleft}

The authors thank Prof. Lin Li from huazhong university of science
and technology and Prof. Yuan Sun from national university of defense
technology for valued discussions and critical reading our manuscript.
This work was supported by National Key Research and Development Program
of China (2017YFA0304800), the National Natural Science Foundation
of China (Grant Nos. 61525504, 61722510, 61435011, 11174271, 61275115,
11604322), Anhui Initiative in Quantum Information Technologies (AHY020200),
and the Youth Innovation Pro motion Association of Chinese Academy
of Sciences under Grant No. 2018490.


\begin{thebibliography}{60}%
\makeatletter
\providecommand \@ifxundefined [1]{%
 \@ifx{#1\undefined}
}%
\providecommand \@ifnum [1]{%
 \ifnum #1\expandafter \@firstoftwo
 \else \expandafter \@secondoftwo
 \fi
}%
\providecommand \@ifx [1]{%
 \ifx #1\expandafter \@firstoftwo
 \else \expandafter \@secondoftwo
 \fi
}%
\providecommand \natexlab [1]{#1}%
\providecommand \enquote  [1]{``#1''}%
\providecommand \bibnamefont  [1]{#1}%
\providecommand \bibfnamefont [1]{#1}%
\providecommand \citenamefont [1]{#1}%
\providecommand \href@noop [0]{\@secondoftwo}%
\providecommand \href [0]{\begingroup \@sanitize@url \@href}%
\providecommand \@href[1]{\@@startlink{#1}\@@href}%
\providecommand \@@href[1]{\endgroup#1\@@endlink}%
\providecommand \@sanitize@url [0]{\catcode `\\12\catcode `\$12\catcode
  `\&12\catcode `\#12\catcode `\^12\catcode `\_12\catcode `\%12\relax}%
\providecommand \@@startlink[1]{}%
\providecommand \@@endlink[0]{}%
\providecommand \url  [0]{\begingroup\@sanitize@url \@url }%
\providecommand \@url [1]{\endgroup\@href {#1}{\urlprefix }}%
\providecommand \urlprefix  [0]{URL }%
\providecommand \Eprint [0]{\href }%
\providecommand \doibase [0]{http://dx.doi.org/}%
\providecommand \selectlanguage [0]{\@gobble}%
\providecommand \bibinfo  [0]{\@secondoftwo}%
\providecommand \bibfield  [0]{\@secondoftwo}%
\providecommand \translation [1]{[#1]}%
\providecommand \BibitemOpen [0]{}%
\providecommand \bibitemStop [0]{}%
\providecommand \bibitemNoStop [0]{.\EOS\space}%
\providecommand \EOS [0]{\spacefactor3000\relax}%
\providecommand \BibitemShut  [1]{\csname bibitem#1\endcsname}%
\let\auto@bib@innerbib\@empty
\bibitem [{\citenamefont {Cirac}\ \emph {et~al.}(1997)\citenamefont {Cirac},
  \citenamefont {Zoller}, \citenamefont {Kimble},\ and\ \citenamefont
  {Mabuchi}}]{cirac1997quantum}%
  \BibitemOpen
  \bibfield  {author} {\bibinfo {author} {\bibfnamefont {Juan~Ignacio}\
  \bibnamefont {Cirac}}, \bibinfo {author} {\bibfnamefont {Peter}\ \bibnamefont
  {Zoller}}, \bibinfo {author} {\bibfnamefont {H~Jeff}\ \bibnamefont {Kimble}},
  \ and\ \bibinfo {author} {\bibfnamefont {Hideo}\ \bibnamefont {Mabuchi}},\
  }\bibfield  {title} {\enquote {\bibinfo {title} {Quantum state transfer and
  entanglement distribution among distant nodes in a quantum network},}\ }\href
  {\doibase 10.1103/PhysRevLett.78.3221} {\bibfield  {journal} {\bibinfo
  {journal} {Phys. Rev. Lett.}\ }\textbf {\bibinfo {volume} {78}},\ \bibinfo
  {pages} {3221} (\bibinfo {year} {1997})}\BibitemShut {NoStop}%
\bibitem [{\citenamefont {O'brien}\ \emph {et~al.}(2009)\citenamefont
  {O'brien}, \citenamefont {Furusawa},\ and\ \citenamefont
  {Vu{\v{c}}kovi{\'c}}}]{o2009photonic}%
  \BibitemOpen
  \bibfield  {author} {\bibinfo {author} {\bibfnamefont {Jeremy~L}\
  \bibnamefont {O'brien}}, \bibinfo {author} {\bibfnamefont {Akira}\
  \bibnamefont {Furusawa}}, \ and\ \bibinfo {author} {\bibfnamefont {Jelena}\
  \bibnamefont {Vu{\v{c}}kovi{\'c}}},\ }\bibfield  {title} {\enquote {\bibinfo
  {title} {Photonic quantum technologies},}\ }\href {\doibase
  https://doi.org/10.1038/nphoton.2009.229} {\bibfield  {journal} {\bibinfo
  {journal} {Nat. Photon.}\ }\textbf {\bibinfo {volume} {3}},\ \bibinfo {pages}
  {687} (\bibinfo {year} {2009})}\BibitemShut {NoStop}%
\bibitem [{\citenamefont {Caulfield}\ and\ \citenamefont
  {Dolev}(2010)}]{caulfield2010future}%
  \BibitemOpen
  \bibfield  {author} {\bibinfo {author} {\bibfnamefont {H~John}\ \bibnamefont
  {Caulfield}}\ and\ \bibinfo {author} {\bibfnamefont {Shlomi}\ \bibnamefont
  {Dolev}},\ }\bibfield  {title} {\enquote {\bibinfo {title} {Why future
  supercomputing requires optics},}\ }\href {\doibase
  https://doi.org/10.1038/nphoton.2010.94} {\bibfield  {journal} {\bibinfo
  {journal} {Nat. Photon.}\ }\textbf {\bibinfo {volume} {4}},\ \bibinfo {pages}
  {261} (\bibinfo {year} {2010})}\BibitemShut {NoStop}%
\bibitem [{\citenamefont {Wang}\ \emph {et~al.}(2018)\citenamefont {Wang},
  \citenamefont {Luo}, \citenamefont {Xu}, \citenamefont {Chen},\ and\
  \citenamefont {Yang}}]{Wang_2018}%
  \BibitemOpen
  \bibfield  {author} {\bibinfo {author} {\bibfnamefont {Feng}\ \bibnamefont
  {Wang}}, \bibinfo {author} {\bibfnamefont {Ming-Xing}\ \bibnamefont {Luo}},
  \bibinfo {author} {\bibfnamefont {Gang}\ \bibnamefont {Xu}}, \bibinfo
  {author} {\bibfnamefont {Xiu-Bo}\ \bibnamefont {Chen}}, \ and\ \bibinfo
  {author} {\bibfnamefont {Yi-Xian}\ \bibnamefont {Yang}},\ }\bibfield  {title}
  {\enquote {\bibinfo {title} {Photonic quantum network transmission assisted
  by the weak cross-kerr nonlinearity},}\ }\href {\doibase
  10.1007/s11433-017-9143-y} {\bibfield  {journal} {\bibinfo  {journal}
  {Science China Physics, Mechanics \& Astronomy}\ }\textbf {\bibinfo {volume}
  {61}} (\bibinfo {year} {2018}),\ 10.1007/s11433-017-9143-y}\BibitemShut
  {NoStop}%
\bibitem [{\citenamefont {Saffman}\ \emph {et~al.}(2010)\citenamefont
  {Saffman}, \citenamefont {Walker},\ and\ \citenamefont
  {M{\o}lmer}}]{saffman2010quantum}%
  \BibitemOpen
  \bibfield  {author} {\bibinfo {author} {\bibfnamefont {Mark}\ \bibnamefont
  {Saffman}}, \bibinfo {author} {\bibfnamefont {TG}~\bibnamefont {Walker}}, \
  and\ \bibinfo {author} {\bibfnamefont {Klaus}\ \bibnamefont {M{\o}lmer}},\
  }\bibfield  {title} {\enquote {\bibinfo {title} {Quantum information with
  rydberg atoms},}\ }\href {\doibase 10.1103/RevModPhys.82.2313} {\bibfield
  {journal} {\bibinfo  {journal} {Reviews of Modern Physics}\ }\textbf
  {\bibinfo {volume} {82}},\ \bibinfo {pages} {2313} (\bibinfo {year}
  {2010})}\BibitemShut {NoStop}%
\bibitem [{\citenamefont {Kimble}(2008)}]{kimble2008quantum}%
  \BibitemOpen
  \bibfield  {author} {\bibinfo {author} {\bibfnamefont {H~Jeff}\ \bibnamefont
  {Kimble}},\ }\bibfield  {title} {\enquote {\bibinfo {title} {The quantum
  internet},}\ }\href {\doibase https://doi.org/10.1038/nature07127} {\bibfield
   {journal} {\bibinfo  {journal} {Nature}\ }\textbf {\bibinfo {volume}
  {453}},\ \bibinfo {pages} {1023} (\bibinfo {year} {2008})}\BibitemShut
  {NoStop}%
\bibitem [{\citenamefont {Cui}\ \emph {et~al.}(2019)\citenamefont {Cui},
  \citenamefont {Zhong}, \citenamefont {Zhou},\ and\ \citenamefont
  {Sheng}}]{Cui_2019}%
  \BibitemOpen
  \bibfield  {author} {\bibinfo {author} {\bibfnamefont {Zheng-Xia}\
  \bibnamefont {Cui}}, \bibinfo {author} {\bibfnamefont {Wei}\ \bibnamefont
  {Zhong}}, \bibinfo {author} {\bibfnamefont {Lan}\ \bibnamefont {Zhou}}, \
  and\ \bibinfo {author} {\bibfnamefont {Yu-Bo}\ \bibnamefont {Sheng}},\
  }\bibfield  {title} {\enquote {\bibinfo {title}
  {Measurement-device-independent quantum key distribution with
  hyper-encoding},}\ }\href {\doibase 10.1007/s11433-019-1438-6} {\bibfield
  {journal} {\bibinfo  {journal} {Science China Physics, Mechanics {\&}
  Astronomy}\ }\textbf {\bibinfo {volume} {62}} (\bibinfo {year} {2019}),\
  10.1007/s11433-019-1438-6}\BibitemShut {NoStop}%
\bibitem [{\citenamefont {Xie}(2020)}]{Xie_2020}%
  \BibitemOpen
  \bibfield  {author} {\bibinfo {author} {\bibfnamefont {XinCheng}\
  \bibnamefont {Xie}},\ }\bibfield  {title} {\enquote {\bibinfo {title}
  {Quantum secure direct communication with an untrusted charlie using
  imperfect measurement devices},}\ }\href {\doibase 10.1007/s11433-019-1491-1}
  {\bibfield  {journal} {\bibinfo  {journal} {Science China Physics, Mechanics
  {\&} Astronomy}\ }\textbf {\bibinfo {volume} {63}} (\bibinfo {year} {2020}),\
  10.1007/s11433-019-1491-1}\BibitemShut {NoStop}%
\bibitem [{\citenamefont {Komar}\ \emph {et~al.}(2014)\citenamefont {Komar},
  \citenamefont {Kessler}, \citenamefont {Bishof}, \citenamefont {Jiang},
  \citenamefont {S{\o}rensen}, \citenamefont {Ye},\ and\ \citenamefont
  {Lukin}}]{komar2014quantum}%
  \BibitemOpen
  \bibfield  {author} {\bibinfo {author} {\bibfnamefont {Peter}\ \bibnamefont
  {Komar}}, \bibinfo {author} {\bibfnamefont {Eric~M}\ \bibnamefont {Kessler}},
  \bibinfo {author} {\bibfnamefont {Michael}\ \bibnamefont {Bishof}}, \bibinfo
  {author} {\bibfnamefont {Liang}\ \bibnamefont {Jiang}}, \bibinfo {author}
  {\bibfnamefont {Anders~S}\ \bibnamefont {S{\o}rensen}}, \bibinfo {author}
  {\bibfnamefont {Jun}\ \bibnamefont {Ye}}, \ and\ \bibinfo {author}
  {\bibfnamefont {Mikhail~D}\ \bibnamefont {Lukin}},\ }\bibfield  {title}
  {\enquote {\bibinfo {title} {A quantum network of clocks},}\ }\href {\doibase
  10.1038/nphys3000} {\bibfield  {journal} {\bibinfo  {journal} {Nat. Phys.}\
  }\textbf {\bibinfo {volume} {10}},\ \bibinfo {pages} {582} (\bibinfo {year}
  {2014})}\BibitemShut {NoStop}%
\bibitem [{\citenamefont {O'Shea}\ \emph {et~al.}(2013)\citenamefont {O'Shea},
  \citenamefont {Junge}, \citenamefont {Volz},\ and\ \citenamefont
  {Rauschenbeutel}}]{o2013fiber}%
  \BibitemOpen
  \bibfield  {author} {\bibinfo {author} {\bibfnamefont {Danny}\ \bibnamefont
  {O'Shea}}, \bibinfo {author} {\bibfnamefont {Christian}\ \bibnamefont
  {Junge}}, \bibinfo {author} {\bibfnamefont {J{\"u}rgen}\ \bibnamefont
  {Volz}}, \ and\ \bibinfo {author} {\bibfnamefont {Arno}\ \bibnamefont
  {Rauschenbeutel}},\ }\bibfield  {title} {\enquote {\bibinfo {title}
  {Fiber-optical switch controlled by a single atom},}\ }\href {\doibase
  10.1103/PhysRevLett.111.193601} {\bibfield  {journal} {\bibinfo  {journal}
  {Phys. Rev. Lett.}\ }\textbf {\bibinfo {volume} {111}},\ \bibinfo {pages}
  {193601} (\bibinfo {year} {2013})}\BibitemShut {NoStop}%
\bibitem [{\citenamefont {Bajcsy}\ \emph {et~al.}(2009)\citenamefont {Bajcsy},
  \citenamefont {Hofferberth}, \citenamefont {Balic}, \citenamefont {Peyronel},
  \citenamefont {Hafezi}, \citenamefont {Zibrov}, \citenamefont {Vuletic},\
  and\ \citenamefont {Lukin}}]{bajcsy2009efficient}%
  \BibitemOpen
  \bibfield  {author} {\bibinfo {author} {\bibfnamefont {Michal}\ \bibnamefont
  {Bajcsy}}, \bibinfo {author} {\bibfnamefont {Sebastian}\ \bibnamefont
  {Hofferberth}}, \bibinfo {author} {\bibfnamefont {Vlatko}\ \bibnamefont
  {Balic}}, \bibinfo {author} {\bibfnamefont {Thibault}\ \bibnamefont
  {Peyronel}}, \bibinfo {author} {\bibfnamefont {Mohammad}\ \bibnamefont
  {Hafezi}}, \bibinfo {author} {\bibfnamefont {Alexander~S}\ \bibnamefont
  {Zibrov}}, \bibinfo {author} {\bibfnamefont {Vladan}\ \bibnamefont
  {Vuletic}}, \ and\ \bibinfo {author} {\bibfnamefont {Mikhail~D}\ \bibnamefont
  {Lukin}},\ }\bibfield  {title} {\enquote {\bibinfo {title} {Efficient
  all-optical switching using slow light within a hollow fiber},}\ }\href
  {\doibase 10.1103/PhysRevLett.102.203902} {\bibfield  {journal} {\bibinfo
  {journal} {Phys. Rev. Lett.}\ }\textbf {\bibinfo {volume} {102}},\ \bibinfo
  {pages} {203902} (\bibinfo {year} {2009})}\BibitemShut {NoStop}%
\bibitem [{\citenamefont {Chen}\ \emph {et~al.}(2013)\citenamefont {Chen},
  \citenamefont {Beck}, \citenamefont {B{\"u}cker}, \citenamefont {Gullans},
  \citenamefont {Lukin}, \citenamefont {Tanji-Suzuki},\ and\ \citenamefont
  {Vuleti{\'c}}}]{chen2013all}%
  \BibitemOpen
  \bibfield  {author} {\bibinfo {author} {\bibfnamefont {Wenlan}\ \bibnamefont
  {Chen}}, \bibinfo {author} {\bibfnamefont {Kristin~M}\ \bibnamefont {Beck}},
  \bibinfo {author} {\bibfnamefont {Robert}\ \bibnamefont {B{\"u}cker}},
  \bibinfo {author} {\bibfnamefont {Michael}\ \bibnamefont {Gullans}}, \bibinfo
  {author} {\bibfnamefont {Mikhail~D}\ \bibnamefont {Lukin}}, \bibinfo {author}
  {\bibfnamefont {Haruka}\ \bibnamefont {Tanji-Suzuki}}, \ and\ \bibinfo
  {author} {\bibfnamefont {Vladan}\ \bibnamefont {Vuleti{\'c}}},\ }\bibfield
  {title} {\enquote {\bibinfo {title} {All-optical switch and transistor gated
  by one stored photon},}\ }\href {\doibase 10.1126/science.1238169} {\bibfield
   {journal} {\bibinfo  {journal} {Science}\ ,\ \bibinfo {pages} {1237242}}
  (\bibinfo {year} {2013})}\BibitemShut {NoStop}%
\bibitem [{\citenamefont {Volz}\ \emph {et~al.}(2012)\citenamefont {Volz},
  \citenamefont {Reinhard}, \citenamefont {Winger}, \citenamefont {Badolato},
  \citenamefont {Hennessy}, \citenamefont {Hu},\ and\ \citenamefont
  {Imamo{\u{g}}lu}}]{volz2012ultrafast}%
  \BibitemOpen
  \bibfield  {author} {\bibinfo {author} {\bibfnamefont {Thomas}\ \bibnamefont
  {Volz}}, \bibinfo {author} {\bibfnamefont {Andreas}\ \bibnamefont
  {Reinhard}}, \bibinfo {author} {\bibfnamefont {Martin}\ \bibnamefont
  {Winger}}, \bibinfo {author} {\bibfnamefont {Antonio}\ \bibnamefont
  {Badolato}}, \bibinfo {author} {\bibfnamefont {Kevin~J}\ \bibnamefont
  {Hennessy}}, \bibinfo {author} {\bibfnamefont {Evelyn~L}\ \bibnamefont {Hu}},
  \ and\ \bibinfo {author} {\bibfnamefont {Ata{\c{c}}}\ \bibnamefont
  {Imamo{\u{g}}lu}},\ }\bibfield  {title} {\enquote {\bibinfo {title}
  {Ultrafast all-optical switching by single photons},}\ }\href {\doibase
  10.1038/nphoton.2012.181} {\bibfield  {journal} {\bibinfo  {journal} {Nat.
  Photon.}\ }\textbf {\bibinfo {volume} {6}},\ \bibinfo {pages} {605} (\bibinfo
  {year} {2012})}\BibitemShut {NoStop}%
\bibitem [{\citenamefont {Hwang}\ \emph {et~al.}(2009)\citenamefont {Hwang},
  \citenamefont {Pototschnig}, \citenamefont {Lettow}, \citenamefont {Zumofen},
  \citenamefont {Renn}, \citenamefont {G{\"o}tzinger},\ and\ \citenamefont
  {Sandoghdar}}]{hwang2009single}%
  \BibitemOpen
  \bibfield  {author} {\bibinfo {author} {\bibfnamefont {Jaesuk}\ \bibnamefont
  {Hwang}}, \bibinfo {author} {\bibfnamefont {Martin}\ \bibnamefont
  {Pototschnig}}, \bibinfo {author} {\bibfnamefont {Robert}\ \bibnamefont
  {Lettow}}, \bibinfo {author} {\bibfnamefont {Gert}\ \bibnamefont {Zumofen}},
  \bibinfo {author} {\bibfnamefont {Alois}\ \bibnamefont {Renn}}, \bibinfo
  {author} {\bibfnamefont {Stephan}\ \bibnamefont {G{\"o}tzinger}}, \ and\
  \bibinfo {author} {\bibfnamefont {Vahid}\ \bibnamefont {Sandoghdar}},\
  }\bibfield  {title} {\enquote {\bibinfo {title} {A single-molecule optical
  transistor},}\ }\href {\doibase https://doi.org/10.1038/nature08134}
  {\bibfield  {journal} {\bibinfo  {journal} {Nature}\ }\textbf {\bibinfo
  {volume} {460}},\ \bibinfo {pages} {76} (\bibinfo {year} {2009})}\BibitemShut
  {NoStop}%
\bibitem [{\citenamefont {Comparat}\ and\ \citenamefont
  {Pillet}(2010)}]{comparat2010dipole}%
  \BibitemOpen
  \bibfield  {author} {\bibinfo {author} {\bibfnamefont {Daniel}\ \bibnamefont
  {Comparat}}\ and\ \bibinfo {author} {\bibfnamefont {Pierre}\ \bibnamefont
  {Pillet}},\ }\bibfield  {title} {\enquote {\bibinfo {title} {Dipole blockade
  in a cold rydberg atomic sample [invited]},}\ }\href {\doibase
  https://doi.org/10.1364/JOSAB.27.00A208} {\bibfield  {journal} {\bibinfo
  {journal} {JOSA B}\ }\textbf {\bibinfo {volume} {27}},\ \bibinfo {pages}
  {A208--A232} (\bibinfo {year} {2010})}\BibitemShut {NoStop}%
\bibitem [{\citenamefont {Jaksch}\ \emph {et~al.}(2000)\citenamefont {Jaksch},
  \citenamefont {Cirac}, \citenamefont {Zoller}, \citenamefont {Rolston},
  \citenamefont {C{\^o}t{\'e}},\ and\ \citenamefont {Lukin}}]{jaksch2000fast}%
  \BibitemOpen
  \bibfield  {author} {\bibinfo {author} {\bibfnamefont {D}~\bibnamefont
  {Jaksch}}, \bibinfo {author} {\bibfnamefont {JI}~\bibnamefont {Cirac}},
  \bibinfo {author} {\bibfnamefont {P}~\bibnamefont {Zoller}}, \bibinfo
  {author} {\bibfnamefont {SL}~\bibnamefont {Rolston}}, \bibinfo {author}
  {\bibfnamefont {R}~\bibnamefont {C{\^o}t{\'e}}}, \ and\ \bibinfo {author}
  {\bibfnamefont {MD}~\bibnamefont {Lukin}},\ }\bibfield  {title} {\enquote
  {\bibinfo {title} {Fast quantum gates for neutral atoms},}\ }\href {\doibase
  10.1103/PhysRevLett.85.2208} {\bibfield  {journal} {\bibinfo  {journal}
  {Phys. Rev. Lett.}\ }\textbf {\bibinfo {volume} {85}},\ \bibinfo {pages}
  {2208} (\bibinfo {year} {2000})}\BibitemShut {NoStop}%
\bibitem [{\citenamefont {Lukin}\ \emph {et~al.}(2001)\citenamefont {Lukin},
  \citenamefont {Fleischhauer}, \citenamefont {Cote}, \citenamefont {Duan},
  \citenamefont {Jaksch}, \citenamefont {Cirac},\ and\ \citenamefont
  {Zoller}}]{lukin2001dipole}%
  \BibitemOpen
  \bibfield  {author} {\bibinfo {author} {\bibfnamefont {MD}~\bibnamefont
  {Lukin}}, \bibinfo {author} {\bibfnamefont {M}~\bibnamefont {Fleischhauer}},
  \bibinfo {author} {\bibfnamefont {R}~\bibnamefont {Cote}}, \bibinfo {author}
  {\bibfnamefont {LM}~\bibnamefont {Duan}}, \bibinfo {author} {\bibfnamefont
  {D}~\bibnamefont {Jaksch}}, \bibinfo {author} {\bibfnamefont
  {JI}~\bibnamefont {Cirac}}, \ and\ \bibinfo {author} {\bibfnamefont
  {P}~\bibnamefont {Zoller}},\ }\bibfield  {title} {\enquote {\bibinfo {title}
  {Dipole blockade and quantum information processing in mesoscopic atomic
  ensembles},}\ }\href {\doibase 10.1103/PhysRevLett.87.037901} {\bibfield
  {journal} {\bibinfo  {journal} {Phys. Rev. Lett.}\ }\textbf {\bibinfo
  {volume} {87}},\ \bibinfo {pages} {037901} (\bibinfo {year}
  {2001})}\BibitemShut {NoStop}%
\bibitem [{\citenamefont {Tong}\ \emph {et~al.}(2004)\citenamefont {Tong},
  \citenamefont {Farooqi}, \citenamefont {Stanojevic}, \citenamefont
  {Krishnan}, \citenamefont {Zhang}, \citenamefont {C{\^o}t{\'e}},
  \citenamefont {Eyler},\ and\ \citenamefont {Gould}}]{tong2004local}%
  \BibitemOpen
  \bibfield  {author} {\bibinfo {author} {\bibfnamefont {D}~\bibnamefont
  {Tong}}, \bibinfo {author} {\bibfnamefont {SM}~\bibnamefont {Farooqi}},
  \bibinfo {author} {\bibfnamefont {J}~\bibnamefont {Stanojevic}}, \bibinfo
  {author} {\bibfnamefont {S}~\bibnamefont {Krishnan}}, \bibinfo {author}
  {\bibfnamefont {YP}~\bibnamefont {Zhang}}, \bibinfo {author} {\bibfnamefont
  {R}~\bibnamefont {C{\^o}t{\'e}}}, \bibinfo {author} {\bibfnamefont
  {EE}~\bibnamefont {Eyler}}, \ and\ \bibinfo {author} {\bibfnamefont
  {PL}~\bibnamefont {Gould}},\ }\bibfield  {title} {\enquote {\bibinfo {title}
  {Local blockade of rydberg excitation in an ultracold gas},}\ }\href
  {\doibase 10.1103/PhysRevLett.93.063001} {\bibfield  {journal} {\bibinfo
  {journal} {Phys. Rev. Lett.}\ }\textbf {\bibinfo {volume} {93}},\ \bibinfo
  {pages} {063001} (\bibinfo {year} {2004})}\BibitemShut {NoStop}%
\bibitem [{\citenamefont {Singer}\ \emph {et~al.}(2004)\citenamefont {Singer},
  \citenamefont {Reetz-Lamour}, \citenamefont {Amthor}, \citenamefont
  {Marcassa},\ and\ \citenamefont {Weidem{\"u}ller}}]{singer2004suppression}%
  \BibitemOpen
  \bibfield  {author} {\bibinfo {author} {\bibfnamefont {Kilian}\ \bibnamefont
  {Singer}}, \bibinfo {author} {\bibfnamefont {Markus}\ \bibnamefont
  {Reetz-Lamour}}, \bibinfo {author} {\bibfnamefont {Thomas}\ \bibnamefont
  {Amthor}}, \bibinfo {author} {\bibfnamefont {Luis~Gustavo}\ \bibnamefont
  {Marcassa}}, \ and\ \bibinfo {author} {\bibfnamefont {Matthias}\ \bibnamefont
  {Weidem{\"u}ller}},\ }\bibfield  {title} {\enquote {\bibinfo {title}
  {Suppression of excitation and spectral broadening induced by interactions in
  a cold gas of rydberg atoms},}\ }\href {\doibase
  https://doi.org/10.1103/PhysRevLett.93.163001} {\bibfield  {journal}
  {\bibinfo  {journal} {Phys. Rev. Lett.}\ }\textbf {\bibinfo {volume} {93}},\
  \bibinfo {pages} {163001} (\bibinfo {year} {2004})}\BibitemShut {NoStop}%
\bibitem [{\citenamefont {Urban}\ \emph {et~al.}(2009)\citenamefont {Urban},
  \citenamefont {Johnson}, \citenamefont {Henage}, \citenamefont {Isenhower},
  \citenamefont {Yavuz}, \citenamefont {Walker},\ and\ \citenamefont
  {Saffman}}]{urban2009observation}%
  \BibitemOpen
  \bibfield  {author} {\bibinfo {author} {\bibfnamefont {E}~\bibnamefont
  {Urban}}, \bibinfo {author} {\bibfnamefont {Todd~A}\ \bibnamefont {Johnson}},
  \bibinfo {author} {\bibfnamefont {T}~\bibnamefont {Henage}}, \bibinfo
  {author} {\bibfnamefont {L}~\bibnamefont {Isenhower}}, \bibinfo {author}
  {\bibfnamefont {DD}~\bibnamefont {Yavuz}}, \bibinfo {author} {\bibfnamefont
  {TG}~\bibnamefont {Walker}}, \ and\ \bibinfo {author} {\bibfnamefont
  {M}~\bibnamefont {Saffman}},\ }\bibfield  {title} {\enquote {\bibinfo {title}
  {Observation of rydberg blockade between two atoms},}\ }\href {\doibase
  10.1038/nphys1178} {\bibfield  {journal} {\bibinfo  {journal} {Nature
  Physics}\ }\textbf {\bibinfo {volume} {5}},\ \bibinfo {pages} {110} (\bibinfo
  {year} {2009})}\BibitemShut {NoStop}%
\bibitem [{\citenamefont {Ga{\"e}tan}\ \emph {et~al.}(2009)\citenamefont
  {Ga{\"e}tan}, \citenamefont {Miroshnychenko}, \citenamefont {Wilk},
  \citenamefont {Chotia}, \citenamefont {Viteau}, \citenamefont {Comparat},
  \citenamefont {Pillet}, \citenamefont {Browaeys},\ and\ \citenamefont
  {Grangier}}]{gaetan2009observation}%
  \BibitemOpen
  \bibfield  {author} {\bibinfo {author} {\bibfnamefont {Alpha}\ \bibnamefont
  {Ga{\"e}tan}}, \bibinfo {author} {\bibfnamefont {Yevhen}\ \bibnamefont
  {Miroshnychenko}}, \bibinfo {author} {\bibfnamefont {Tatjana}\ \bibnamefont
  {Wilk}}, \bibinfo {author} {\bibfnamefont {Amodsen}\ \bibnamefont {Chotia}},
  \bibinfo {author} {\bibfnamefont {Matthieu}\ \bibnamefont {Viteau}}, \bibinfo
  {author} {\bibfnamefont {Daniel}\ \bibnamefont {Comparat}}, \bibinfo {author}
  {\bibfnamefont {Pierre}\ \bibnamefont {Pillet}}, \bibinfo {author}
  {\bibfnamefont {Antoine}\ \bibnamefont {Browaeys}}, \ and\ \bibinfo {author}
  {\bibfnamefont {Philippe}\ \bibnamefont {Grangier}},\ }\bibfield  {title}
  {\enquote {\bibinfo {title} {Observation of collective excitation of two
  individual atoms in the rydberg blockade regime},}\ }\href {\doibase
  10.1038/nphys1183} {\bibfield  {journal} {\bibinfo  {journal} {Nat. Phys.}\
  }\textbf {\bibinfo {volume} {5}},\ \bibinfo {pages} {115} (\bibinfo {year}
  {2009})}\BibitemShut {NoStop}%
\bibitem [{\citenamefont {Heidemann}\ \emph {et~al.}(2007)\citenamefont
  {Heidemann}, \citenamefont {Raitzsch}, \citenamefont {Bendkowsky},
  \citenamefont {Butscher}, \citenamefont {L{\"o}w}, \citenamefont {Santos},\
  and\ \citenamefont {Pfau}}]{heidemann2007evidence}%
  \BibitemOpen
  \bibfield  {author} {\bibinfo {author} {\bibfnamefont {Rolf}\ \bibnamefont
  {Heidemann}}, \bibinfo {author} {\bibfnamefont {Ulrich}\ \bibnamefont
  {Raitzsch}}, \bibinfo {author} {\bibfnamefont {Vera}\ \bibnamefont
  {Bendkowsky}}, \bibinfo {author} {\bibfnamefont {Bj{\"o}rn}\ \bibnamefont
  {Butscher}}, \bibinfo {author} {\bibfnamefont {Robert}\ \bibnamefont
  {L{\"o}w}}, \bibinfo {author} {\bibfnamefont {Luis}\ \bibnamefont {Santos}},
  \ and\ \bibinfo {author} {\bibfnamefont {Tilman}\ \bibnamefont {Pfau}},\
  }\bibfield  {title} {\enquote {\bibinfo {title} {Evidence for coherent
  collective rydberg excitation in the strong blockade regime},}\ }\href
  {\doibase https://doi.org/10.1103/PhysRevLett.99.163601} {\bibfield
  {journal} {\bibinfo  {journal} {Phys. Rev. Lett.}\ }\textbf {\bibinfo
  {volume} {99}},\ \bibinfo {pages} {163601} (\bibinfo {year}
  {2007})}\BibitemShut {NoStop}%
\bibitem [{\citenamefont {Zeiher}\ \emph {et~al.}(2015)\citenamefont {Zeiher},
  \citenamefont {Schau{\ss}}, \citenamefont {Hild}, \citenamefont {Macr{\`\i}},
  \citenamefont {Bloch},\ and\ \citenamefont {Gross}}]{zeiher2015microscopic}%
  \BibitemOpen
  \bibfield  {author} {\bibinfo {author} {\bibfnamefont {Johannes}\
  \bibnamefont {Zeiher}}, \bibinfo {author} {\bibfnamefont {Peter}\
  \bibnamefont {Schau{\ss}}}, \bibinfo {author} {\bibfnamefont {Sebastian}\
  \bibnamefont {Hild}}, \bibinfo {author} {\bibfnamefont {Tommaso}\
  \bibnamefont {Macr{\`\i}}}, \bibinfo {author} {\bibfnamefont {Immanuel}\
  \bibnamefont {Bloch}}, \ and\ \bibinfo {author} {\bibfnamefont {Christian}\
  \bibnamefont {Gross}},\ }\bibfield  {title} {\enquote {\bibinfo {title}
  {Microscopic characterization of scalable coherent rydberg superatoms},}\
  }\href {\doibase https://doi.org/10.1103/PhysRevX.5.031015} {\bibfield
  {journal} {\bibinfo  {journal} {Phys. Rev. X}\ }\textbf {\bibinfo {volume}
  {5}},\ \bibinfo {pages} {031015} (\bibinfo {year} {2015})}\BibitemShut
  {NoStop}%
\bibitem [{\citenamefont {Labuhn}\ \emph {et~al.}(2016)\citenamefont {Labuhn},
  \citenamefont {Barredo}, \citenamefont {Ravets}, \citenamefont
  {De~L{\'e}s{\'e}leuc}, \citenamefont {Macr{\`\i}}, \citenamefont {Lahaye},\
  and\ \citenamefont {Browaeys}}]{labuhn2016tunable}%
  \BibitemOpen
  \bibfield  {author} {\bibinfo {author} {\bibfnamefont {Henning}\ \bibnamefont
  {Labuhn}}, \bibinfo {author} {\bibfnamefont {Daniel}\ \bibnamefont
  {Barredo}}, \bibinfo {author} {\bibfnamefont {Sylvain}\ \bibnamefont
  {Ravets}}, \bibinfo {author} {\bibfnamefont {Sylvain}\ \bibnamefont
  {De~L{\'e}s{\'e}leuc}}, \bibinfo {author} {\bibfnamefont {Tommaso}\
  \bibnamefont {Macr{\`\i}}}, \bibinfo {author} {\bibfnamefont {Thierry}\
  \bibnamefont {Lahaye}}, \ and\ \bibinfo {author} {\bibfnamefont {Antoine}\
  \bibnamefont {Browaeys}},\ }\bibfield  {title} {\enquote {\bibinfo {title}
  {Tunable two-dimensional arrays of single rydberg atoms for realizing quantum
  ising models},}\ }\href {\doibase 10.1038/nature18274} {\bibfield  {journal}
  {\bibinfo  {journal} {Nature}\ }\textbf {\bibinfo {volume} {534}},\ \bibinfo
  {pages} {667--684} (\bibinfo {year} {2016})}\BibitemShut {NoStop}%
\bibitem [{\citenamefont {Bernien}\ \emph {et~al.}(2017)\citenamefont
  {Bernien}, \citenamefont {Schwartz}, \citenamefont {Keesling}, \citenamefont
  {Levine}, \citenamefont {Omran}, \citenamefont {Pichler}, \citenamefont
  {Choi}, \citenamefont {Zibrov}, \citenamefont {Endres}, \citenamefont
  {Greiner} \emph {et~al.}}]{bernien2017probing}%
  \BibitemOpen
  \bibfield  {author} {\bibinfo {author} {\bibfnamefont {Hannes}\ \bibnamefont
  {Bernien}}, \bibinfo {author} {\bibfnamefont {Sylvain}\ \bibnamefont
  {Schwartz}}, \bibinfo {author} {\bibfnamefont {Alexander}\ \bibnamefont
  {Keesling}}, \bibinfo {author} {\bibfnamefont {Harry}\ \bibnamefont
  {Levine}}, \bibinfo {author} {\bibfnamefont {Ahmed}\ \bibnamefont {Omran}},
  \bibinfo {author} {\bibfnamefont {Hannes}\ \bibnamefont {Pichler}}, \bibinfo
  {author} {\bibfnamefont {Soonwon}\ \bibnamefont {Choi}}, \bibinfo {author}
  {\bibfnamefont {Alexander~S}\ \bibnamefont {Zibrov}}, \bibinfo {author}
  {\bibfnamefont {Manuel}\ \bibnamefont {Endres}}, \bibinfo {author}
  {\bibfnamefont {Markus}\ \bibnamefont {Greiner}},  \emph {et~al.},\
  }\bibfield  {title} {\enquote {\bibinfo {title} {Probing many-body dynamics
  on a 51-atom quantum simulator},}\ }\href {\doibase 10.1038/nature24622}
  {\bibfield  {journal} {\bibinfo  {journal} {Nature}\ }\textbf {\bibinfo
  {volume} {551}},\ \bibinfo {pages} {579} (\bibinfo {year}
  {2017})}\BibitemShut {NoStop}%
\bibitem [{\citenamefont {Schau{\ss}}\ \emph {et~al.}(2012)\citenamefont
  {Schau{\ss}}, \citenamefont {Cheneau}, \citenamefont {Endres}, \citenamefont
  {Fukuhara}, \citenamefont {Hild}, \citenamefont {Omran}, \citenamefont
  {Pohl}, \citenamefont {Gross}, \citenamefont {Kuhr},\ and\ \citenamefont
  {Bloch}}]{schauss2012observation}%
  \BibitemOpen
  \bibfield  {author} {\bibinfo {author} {\bibfnamefont {Peter}\ \bibnamefont
  {Schau{\ss}}}, \bibinfo {author} {\bibfnamefont {Marc}\ \bibnamefont
  {Cheneau}}, \bibinfo {author} {\bibfnamefont {Manuel}\ \bibnamefont
  {Endres}}, \bibinfo {author} {\bibfnamefont {Takeshi}\ \bibnamefont
  {Fukuhara}}, \bibinfo {author} {\bibfnamefont {Sebastian}\ \bibnamefont
  {Hild}}, \bibinfo {author} {\bibfnamefont {Ahmed}\ \bibnamefont {Omran}},
  \bibinfo {author} {\bibfnamefont {Thomas}\ \bibnamefont {Pohl}}, \bibinfo
  {author} {\bibfnamefont {Christian}\ \bibnamefont {Gross}}, \bibinfo {author}
  {\bibfnamefont {Stefan}\ \bibnamefont {Kuhr}}, \ and\ \bibinfo {author}
  {\bibfnamefont {Immanuel}\ \bibnamefont {Bloch}},\ }\bibfield  {title}
  {\enquote {\bibinfo {title} {Observation of spatially ordered structures in a
  two-dimensional rydberg gas},}\ }\href {\doibase 10.1038/nature11596}
  {\bibfield  {journal} {\bibinfo  {journal} {Nature}\ }\textbf {\bibinfo
  {volume} {491}},\ \bibinfo {pages} {87--91} (\bibinfo {year}
  {2012})}\BibitemShut {NoStop}%
\bibitem [{\citenamefont {Schwarzkopf}\ \emph {et~al.}(2011)\citenamefont
  {Schwarzkopf}, \citenamefont {Sapiro},\ and\ \citenamefont
  {Raithel}}]{schwarzkopf2011imaging}%
  \BibitemOpen
  \bibfield  {author} {\bibinfo {author} {\bibfnamefont {Andrew}\ \bibnamefont
  {Schwarzkopf}}, \bibinfo {author} {\bibfnamefont {RE}~\bibnamefont {Sapiro}},
  \ and\ \bibinfo {author} {\bibfnamefont {Georg}\ \bibnamefont {Raithel}},\
  }\bibfield  {title} {\enquote {\bibinfo {title} {Imaging spatial correlations
  of rydberg excitations in cold atom clouds},}\ }\href {\doibase
  10.1103/PhysRevLett.107.103001} {\bibfield  {journal} {\bibinfo  {journal}
  {Phys. Rev. Lett.}\ }\textbf {\bibinfo {volume} {107}},\ \bibinfo {pages}
  {103001} (\bibinfo {year} {2011})}\BibitemShut {NoStop}%
\bibitem [{\citenamefont {Schau{\ss}}\ \emph {et~al.}(2015)\citenamefont
  {Schau{\ss}}, \citenamefont {Zeiher}, \citenamefont {Fukuhara}, \citenamefont
  {Hild}, \citenamefont {Cheneau}, \citenamefont {Macr{\`\i}}, \citenamefont
  {Pohl}, \citenamefont {Bloch},\ and\ \citenamefont
  {Gro{\ss}}}]{schauss2015crystallization}%
  \BibitemOpen
  \bibfield  {author} {\bibinfo {author} {\bibfnamefont {Peter}\ \bibnamefont
  {Schau{\ss}}}, \bibinfo {author} {\bibfnamefont {Johannes}\ \bibnamefont
  {Zeiher}}, \bibinfo {author} {\bibfnamefont {Takeshi}\ \bibnamefont
  {Fukuhara}}, \bibinfo {author} {\bibfnamefont {Sebastian}\ \bibnamefont
  {Hild}}, \bibinfo {author} {\bibfnamefont {Marc}\ \bibnamefont {Cheneau}},
  \bibinfo {author} {\bibfnamefont {T}~\bibnamefont {Macr{\`\i}}}, \bibinfo
  {author} {\bibfnamefont {Thomas}\ \bibnamefont {Pohl}}, \bibinfo {author}
  {\bibfnamefont {Immanuel}\ \bibnamefont {Bloch}}, \ and\ \bibinfo {author}
  {\bibfnamefont {Christian}\ \bibnamefont {Gro{\ss}}},\ }\bibfield  {title}
  {\enquote {\bibinfo {title} {Crystallization in ising quantum magnets},}\
  }\href {\doibase 10.1126/science.1258351} {\bibfield  {journal} {\bibinfo
  {journal} {Science}\ }\textbf {\bibinfo {volume} {347}},\ \bibinfo {pages}
  {1455--1458} (\bibinfo {year} {2015})}\BibitemShut {NoStop}%
\bibitem [{\citenamefont {Pritchard}\ \emph
  {et~al.}(2010{\natexlab{a}})\citenamefont {Pritchard}, \citenamefont
  {Maxwell}, \citenamefont {Gauguet}, \citenamefont {Weatherill}, \citenamefont
  {Jones},\ and\ \citenamefont {Adams}}]{pritchard2010cooperative}%
  \BibitemOpen
  \bibfield  {author} {\bibinfo {author} {\bibfnamefont {Jonathan~D}\
  \bibnamefont {Pritchard}}, \bibinfo {author} {\bibfnamefont {D}~\bibnamefont
  {Maxwell}}, \bibinfo {author} {\bibfnamefont {Alexandre}\ \bibnamefont
  {Gauguet}}, \bibinfo {author} {\bibfnamefont {Kevin~J}\ \bibnamefont
  {Weatherill}}, \bibinfo {author} {\bibfnamefont {MPA}\ \bibnamefont {Jones}},
  \ and\ \bibinfo {author} {\bibfnamefont {Charles~S}\ \bibnamefont {Adams}},\
  }\bibfield  {title} {\enquote {\bibinfo {title} {Cooperative atom-light
  interaction in a blockaded rydberg ensemble},}\ }\href {\doibase
  10.1103/PhysRevLett.105.193603} {\bibfield  {journal} {\bibinfo  {journal}
  {Phys. Rev. Lett.}\ }\textbf {\bibinfo {volume} {105}},\ \bibinfo {pages}
  {193603} (\bibinfo {year} {2010}{\natexlab{a}})}\BibitemShut {NoStop}%
\bibitem [{\citenamefont {Dudin}\ and\ \citenamefont
  {Kuzmich}(2012)}]{dudin2012strongly}%
  \BibitemOpen
  \bibfield  {author} {\bibinfo {author} {\bibfnamefont {YO}~\bibnamefont
  {Dudin}}\ and\ \bibinfo {author} {\bibfnamefont {A}~\bibnamefont {Kuzmich}},\
  }\bibfield  {title} {\enquote {\bibinfo {title} {Strongly interacting rydberg
  excitations of a cold atomic gas},}\ }\href {\doibase
  10.1126/science.1217901} {\bibfield  {journal} {\bibinfo  {journal}
  {Science}\ }\textbf {\bibinfo {volume} {336}},\ \bibinfo {pages} {887--889}
  (\bibinfo {year} {2012})}\BibitemShut {NoStop}%
\bibitem [{\citenamefont {Peyronel}\ \emph {et~al.}(2012)\citenamefont
  {Peyronel}, \citenamefont {Firstenberg}, \citenamefont {Liang}, \citenamefont
  {Hofferberth}, \citenamefont {Gorshkov}, \citenamefont {Pohl}, \citenamefont
  {Lukin},\ and\ \citenamefont {Vuleti{\'c}}}]{peyronel2012quantum}%
  \BibitemOpen
  \bibfield  {author} {\bibinfo {author} {\bibfnamefont {Thibault}\
  \bibnamefont {Peyronel}}, \bibinfo {author} {\bibfnamefont {Ofer}\
  \bibnamefont {Firstenberg}}, \bibinfo {author} {\bibfnamefont {Qi-Yu}\
  \bibnamefont {Liang}}, \bibinfo {author} {\bibfnamefont {Sebastian}\
  \bibnamefont {Hofferberth}}, \bibinfo {author} {\bibfnamefont {Alexey~V}\
  \bibnamefont {Gorshkov}}, \bibinfo {author} {\bibfnamefont {Thomas}\
  \bibnamefont {Pohl}}, \bibinfo {author} {\bibfnamefont {Mikhail~D}\
  \bibnamefont {Lukin}}, \ and\ \bibinfo {author} {\bibfnamefont {Vladan}\
  \bibnamefont {Vuleti{\'c}}},\ }\bibfield  {title} {\enquote {\bibinfo {title}
  {Quantum nonlinear optics with single photons enabled by strongly interacting
  atoms},}\ }\href {\doibase 10.1038/nature11361} {\bibfield  {journal}
  {\bibinfo  {journal} {Nature}\ }\textbf {\bibinfo {volume} {488}},\ \bibinfo
  {pages} {57--60} (\bibinfo {year} {2012})}\BibitemShut {NoStop}%
\bibitem [{\citenamefont {Maxwell}\ \emph {et~al.}(2013)\citenamefont
  {Maxwell}, \citenamefont {Szwer}, \citenamefont {Paredes-Barato},
  \citenamefont {Busche}, \citenamefont {Pritchard}, \citenamefont {Gauguet},
  \citenamefont {Weatherill}, \citenamefont {Jones},\ and\ \citenamefont
  {Adams}}]{maxwell2013storage}%
  \BibitemOpen
  \bibfield  {author} {\bibinfo {author} {\bibfnamefont {D}~\bibnamefont
  {Maxwell}}, \bibinfo {author} {\bibfnamefont {DJ}~\bibnamefont {Szwer}},
  \bibinfo {author} {\bibfnamefont {D}~\bibnamefont {Paredes-Barato}}, \bibinfo
  {author} {\bibfnamefont {H}~\bibnamefont {Busche}}, \bibinfo {author}
  {\bibfnamefont {JD}~\bibnamefont {Pritchard}}, \bibinfo {author}
  {\bibfnamefont {Alexandre}\ \bibnamefont {Gauguet}}, \bibinfo {author}
  {\bibfnamefont {KJ}~\bibnamefont {Weatherill}}, \bibinfo {author}
  {\bibfnamefont {MPA}\ \bibnamefont {Jones}}, \ and\ \bibinfo {author}
  {\bibfnamefont {CS}~\bibnamefont {Adams}},\ }\bibfield  {title} {\enquote
  {\bibinfo {title} {Storage and control of optical photons using rydberg
  polaritons},}\ }\href {\doibase
  https://doi.org/10.1103/PhysRevLett.110.103001} {\bibfield  {journal}
  {\bibinfo  {journal} {Phys. Rev. Lett.}\ }\textbf {\bibinfo {volume} {110}},\
  \bibinfo {pages} {103001} (\bibinfo {year} {2013})}\BibitemShut {NoStop}%
\bibitem [{\citenamefont {Tresp}\ \emph {et~al.}(2015)\citenamefont {Tresp},
  \citenamefont {Bienias}, \citenamefont {Weber}, \citenamefont {Gorniaczyk},
  \citenamefont {Mirgorodskiy}, \citenamefont {B{\"u}chler},\ and\
  \citenamefont {Hofferberth}}]{tresp2015dipolar}%
  \BibitemOpen
  \bibfield  {author} {\bibinfo {author} {\bibfnamefont {Christoph}\
  \bibnamefont {Tresp}}, \bibinfo {author} {\bibfnamefont {Przemyslaw}\
  \bibnamefont {Bienias}}, \bibinfo {author} {\bibfnamefont {Sebastian}\
  \bibnamefont {Weber}}, \bibinfo {author} {\bibfnamefont {Hannes}\
  \bibnamefont {Gorniaczyk}}, \bibinfo {author} {\bibfnamefont {Ivan}\
  \bibnamefont {Mirgorodskiy}}, \bibinfo {author} {\bibfnamefont {Hans~Peter}\
  \bibnamefont {B{\"u}chler}}, \ and\ \bibinfo {author} {\bibfnamefont
  {Sebastian}\ \bibnamefont {Hofferberth}},\ }\bibfield  {title} {\enquote
  {\bibinfo {title} {Dipolar dephasing of rydberg d-state polaritons},}\ }\href
  {\doibase 10.1103/PhysRevLett.115.083602} {\bibfield  {journal} {\bibinfo
  {journal} {Phys. Rev. Lett.}\ }\textbf {\bibinfo {volume} {115}},\ \bibinfo
  {pages} {083602} (\bibinfo {year} {2015})}\BibitemShut {NoStop}%
\bibitem [{\citenamefont {Firstenberg}\ \emph {et~al.}(2016)\citenamefont
  {Firstenberg}, \citenamefont {Adams},\ and\ \citenamefont
  {Hofferberth}}]{firstenberg2016nonlinear}%
  \BibitemOpen
  \bibfield  {author} {\bibinfo {author} {\bibfnamefont {Ofer}\ \bibnamefont
  {Firstenberg}}, \bibinfo {author} {\bibfnamefont {Charles~S}\ \bibnamefont
  {Adams}}, \ and\ \bibinfo {author} {\bibfnamefont {Sebastian}\ \bibnamefont
  {Hofferberth}},\ }\bibfield  {title} {\enquote {\bibinfo {title} {Nonlinear
  quantum optics mediated by rydberg interactions},}\ }\href {\doibase
  10.1088/0953-4075/49/15/152003} {\bibfield  {journal} {\bibinfo  {journal}
  {Journal of Physics B: Atomic, Molecular and Optical Physics}\ }\textbf
  {\bibinfo {volume} {49}},\ \bibinfo {pages} {152003} (\bibinfo {year}
  {2016})}\BibitemShut {NoStop}%
\bibitem [{\citenamefont {Murray}\ and\ \citenamefont
  {Pohl}(2017)}]{murray2017coherent}%
  \BibitemOpen
  \bibfield  {author} {\bibinfo {author} {\bibfnamefont {Callum~R}\
  \bibnamefont {Murray}}\ and\ \bibinfo {author} {\bibfnamefont {Thomas}\
  \bibnamefont {Pohl}},\ }\bibfield  {title} {\enquote {\bibinfo {title}
  {Coherent photon manipulation in interacting atomic ensembles},}\ }\href
  {\doibase https://doi.org/10.1103/PhysRevX.7.031007} {\bibfield  {journal}
  {\bibinfo  {journal} {Phys. Rev. X}\ }\textbf {\bibinfo {volume} {7}},\
  \bibinfo {pages} {031007} (\bibinfo {year} {2017})}\BibitemShut {NoStop}%
\bibitem [{\citenamefont {Robert-de Saint-Vincent}\ \emph
  {et~al.}(2013)\citenamefont {Robert-de Saint-Vincent}, \citenamefont
  {Hofmann}, \citenamefont {Schempp}, \citenamefont {G{\"u}nter}, \citenamefont
  {Whitlock},\ and\ \citenamefont {Weidem{\"u}ller}}]{robert2013spontaneous}%
  \BibitemOpen
  \bibfield  {author} {\bibinfo {author} {\bibfnamefont {M}~\bibnamefont
  {Robert-de Saint-Vincent}}, \bibinfo {author} {\bibfnamefont
  {CS}~\bibnamefont {Hofmann}}, \bibinfo {author} {\bibfnamefont
  {H}~\bibnamefont {Schempp}}, \bibinfo {author} {\bibfnamefont
  {G}~\bibnamefont {G{\"u}nter}}, \bibinfo {author} {\bibfnamefont
  {S}~\bibnamefont {Whitlock}}, \ and\ \bibinfo {author} {\bibfnamefont
  {M}~\bibnamefont {Weidem{\"u}ller}},\ }\bibfield  {title} {\enquote {\bibinfo
  {title} {Spontaneous avalanche ionization of a strongly blockaded rydberg
  gas},}\ }\href {\doibase https://doi.org/10.1103/PhysRevLett.110.045004}
  {\bibfield  {journal} {\bibinfo  {journal} {Phys. Rev. Lett.}\ }\textbf
  {\bibinfo {volume} {110}},\ \bibinfo {pages} {045004} (\bibinfo {year}
  {2013})}\BibitemShut {NoStop}%
\bibitem [{\citenamefont {Tiarks}\ \emph {et~al.}(2019)\citenamefont {Tiarks},
  \citenamefont {Schmidt-Eberle}, \citenamefont {Stolz}, \citenamefont
  {Rempe},\ and\ \citenamefont {D{\"u}rr}}]{tiarks2019photon}%
  \BibitemOpen
  \bibfield  {author} {\bibinfo {author} {\bibfnamefont {Daniel}\ \bibnamefont
  {Tiarks}}, \bibinfo {author} {\bibfnamefont {Steffen}\ \bibnamefont
  {Schmidt-Eberle}}, \bibinfo {author} {\bibfnamefont {Thomas}\ \bibnamefont
  {Stolz}}, \bibinfo {author} {\bibfnamefont {Gerhard}\ \bibnamefont {Rempe}},
  \ and\ \bibinfo {author} {\bibfnamefont {Stephan}\ \bibnamefont {D{\"u}rr}},\
  }\bibfield  {title} {\enquote {\bibinfo {title} {A photon--photon quantum
  gate based on rydberg interactions},}\ }\href {\doibase
  https://doi.org/10.1038/s41567-018-0313-7} {\bibfield  {journal} {\bibinfo
  {journal} {Nature Physics}\ }\textbf {\bibinfo {volume} {15}},\ \bibinfo
  {pages} {124} (\bibinfo {year} {2019})}\BibitemShut {NoStop}%
\bibitem [{\citenamefont {Tiarks}\ \emph {et~al.}(2014)\citenamefont {Tiarks},
  \citenamefont {Baur}, \citenamefont {Schneider}, \citenamefont {D{\"u}rr},\
  and\ \citenamefont {Rempe}}]{tiarks2014single}%
  \BibitemOpen
  \bibfield  {author} {\bibinfo {author} {\bibfnamefont {Daniel}\ \bibnamefont
  {Tiarks}}, \bibinfo {author} {\bibfnamefont {Simon}\ \bibnamefont {Baur}},
  \bibinfo {author} {\bibfnamefont {Katharina}\ \bibnamefont {Schneider}},
  \bibinfo {author} {\bibfnamefont {Stephan}\ \bibnamefont {D{\"u}rr}}, \ and\
  \bibinfo {author} {\bibfnamefont {Gerhard}\ \bibnamefont {Rempe}},\
  }\bibfield  {title} {\enquote {\bibinfo {title} {Single-photon transistor
  using a f{\"o}rster resonance},}\ }\href {\doibase
  10.1103/PhysRevLett.113.053602} {\bibfield  {journal} {\bibinfo  {journal}
  {Phys. Rev. Lett.}\ }\textbf {\bibinfo {volume} {113}},\ \bibinfo {pages}
  {053602} (\bibinfo {year} {2014})}\BibitemShut {NoStop}%
\bibitem [{\citenamefont {Gorniaczyk}\ \emph {et~al.}(2014)\citenamefont
  {Gorniaczyk}, \citenamefont {Tresp}, \citenamefont {Schmidt}, \citenamefont
  {Fedder},\ and\ \citenamefont {Hofferberth}}]{gorniaczyk2014single}%
  \BibitemOpen
  \bibfield  {author} {\bibinfo {author} {\bibfnamefont {Hannes}\ \bibnamefont
  {Gorniaczyk}}, \bibinfo {author} {\bibfnamefont {Christoph}\ \bibnamefont
  {Tresp}}, \bibinfo {author} {\bibfnamefont {Johannes}\ \bibnamefont
  {Schmidt}}, \bibinfo {author} {\bibfnamefont {Helmut}\ \bibnamefont
  {Fedder}}, \ and\ \bibinfo {author} {\bibfnamefont {Sebastian}\ \bibnamefont
  {Hofferberth}},\ }\bibfield  {title} {\enquote {\bibinfo {title}
  {Single-photon transistor mediated by interstate rydberg interactions},}\
  }\href {\doibase 10.1103/PhysRevLett.113.053601} {\bibfield  {journal}
  {\bibinfo  {journal} {Phys. Rev. Lett.}\ }\textbf {\bibinfo {volume} {113}},\
  \bibinfo {pages} {053601} (\bibinfo {year} {2014})}\BibitemShut {NoStop}%
\bibitem [{\citenamefont {Baur}\ \emph {et~al.}(2014)\citenamefont {Baur},
  \citenamefont {Tiarks}, \citenamefont {Rempe},\ and\ \citenamefont
  {D{\"u}rr}}]{baur2014single}%
  \BibitemOpen
  \bibfield  {author} {\bibinfo {author} {\bibfnamefont {Simon}\ \bibnamefont
  {Baur}}, \bibinfo {author} {\bibfnamefont {Daniel}\ \bibnamefont {Tiarks}},
  \bibinfo {author} {\bibfnamefont {Gerhard}\ \bibnamefont {Rempe}}, \ and\
  \bibinfo {author} {\bibfnamefont {Stephan}\ \bibnamefont {D{\"u}rr}},\
  }\bibfield  {title} {\enquote {\bibinfo {title} {Single-photon switch based
  on rydberg blockade},}\ }\href {\doibase 10.1103/PhysRevLett.112.073901}
  {\bibfield  {journal} {\bibinfo  {journal} {Phys. Rev. Lett.}\ }\textbf
  {\bibinfo {volume} {112}},\ \bibinfo {pages} {073901} (\bibinfo {year}
  {2014})}\BibitemShut {NoStop}%
\bibitem [{\citenamefont {Firstenberg}\ \emph {et~al.}(2013)\citenamefont
  {Firstenberg}, \citenamefont {Peyronel}, \citenamefont {Liang}, \citenamefont
  {Gorshkov}, \citenamefont {Lukin},\ and\ \citenamefont
  {Vuleti{\'c}}}]{firstenberg2013attractive}%
  \BibitemOpen
  \bibfield  {author} {\bibinfo {author} {\bibfnamefont {Ofer}\ \bibnamefont
  {Firstenberg}}, \bibinfo {author} {\bibfnamefont {Thibault}\ \bibnamefont
  {Peyronel}}, \bibinfo {author} {\bibfnamefont {Qi-Yu}\ \bibnamefont {Liang}},
  \bibinfo {author} {\bibfnamefont {Alexey~V}\ \bibnamefont {Gorshkov}},
  \bibinfo {author} {\bibfnamefont {Mikhail~D}\ \bibnamefont {Lukin}}, \ and\
  \bibinfo {author} {\bibfnamefont {Vladan}\ \bibnamefont {Vuleti{\'c}}},\
  }\bibfield  {title} {\enquote {\bibinfo {title} {Attractive photons in a
  quantum nonlinear medium},}\ }\href {\doibase 10.1038/nature12512} {\bibfield
   {journal} {\bibinfo  {journal} {Nature}\ }\textbf {\bibinfo {volume}
  {502}},\ \bibinfo {pages} {71--75} (\bibinfo {year} {2013})}\BibitemShut
  {NoStop}%
\bibitem [{\citenamefont {Petrosyan}\ \emph {et~al.}(2011)\citenamefont
  {Petrosyan}, \citenamefont {Otterbach},\ and\ \citenamefont
  {Fleischhauer}}]{petrosyan2011electromagnetically}%
  \BibitemOpen
  \bibfield  {author} {\bibinfo {author} {\bibfnamefont {David}\ \bibnamefont
  {Petrosyan}}, \bibinfo {author} {\bibfnamefont {Johannes}\ \bibnamefont
  {Otterbach}}, \ and\ \bibinfo {author} {\bibfnamefont {Michael}\ \bibnamefont
  {Fleischhauer}},\ }\bibfield  {title} {\enquote {\bibinfo {title}
  {Electromagnetically induced transparency with rydberg atoms},}\ }\href
  {\doibase 10.1103/physrevlett.107.213601} {\bibfield  {journal} {\bibinfo
  {journal} {Phys. Rev. Lett.}\ }\textbf {\bibinfo {volume} {107}},\ \bibinfo
  {pages} {213601} (\bibinfo {year} {2011})}\BibitemShut {NoStop}%
\bibitem [{\citenamefont {Cory}\ \emph {et~al.}(1998)\citenamefont {Cory},
  \citenamefont {Price}, \citenamefont {Maas}, \citenamefont {Knill},
  \citenamefont {Laflamme}, \citenamefont {Zurek}, \citenamefont {Havel},\ and\
  \citenamefont {Somaroo}}]{cory1998experimental}%
  \BibitemOpen
  \bibfield  {author} {\bibinfo {author} {\bibfnamefont {David~G}\ \bibnamefont
  {Cory}}, \bibinfo {author} {\bibfnamefont {MD}~\bibnamefont {Price}},
  \bibinfo {author} {\bibfnamefont {W}~\bibnamefont {Maas}}, \bibinfo {author}
  {\bibfnamefont {E}~\bibnamefont {Knill}}, \bibinfo {author} {\bibfnamefont
  {Raymond}\ \bibnamefont {Laflamme}}, \bibinfo {author} {\bibfnamefont
  {Wojciech~H}\ \bibnamefont {Zurek}}, \bibinfo {author} {\bibfnamefont
  {Timothy~F}\ \bibnamefont {Havel}}, \ and\ \bibinfo {author} {\bibfnamefont
  {SS}~\bibnamefont {Somaroo}},\ }\bibfield  {title} {\enquote {\bibinfo
  {title} {Experimental quantum error correction},}\ }\href {\doibase
  10.1103/PhysRevLett.81.2152} {\bibfield  {journal} {\bibinfo  {journal}
  {Phys. Rev. Lett.}\ }\textbf {\bibinfo {volume} {81}},\ \bibinfo {pages}
  {2152} (\bibinfo {year} {1998})}\BibitemShut {NoStop}%
\bibitem [{\citenamefont {Gorshkov}\ \emph {et~al.}(2011)\citenamefont
  {Gorshkov}, \citenamefont {Otterbach}, \citenamefont {Fleischhauer},
  \citenamefont {Pohl},\ and\ \citenamefont {Lukin}}]{gorshkov2011photon}%
  \BibitemOpen
  \bibfield  {author} {\bibinfo {author} {\bibfnamefont {Alexey~V}\
  \bibnamefont {Gorshkov}}, \bibinfo {author} {\bibfnamefont {Johannes}\
  \bibnamefont {Otterbach}}, \bibinfo {author} {\bibfnamefont {Michael}\
  \bibnamefont {Fleischhauer}}, \bibinfo {author} {\bibfnamefont {Thomas}\
  \bibnamefont {Pohl}}, \ and\ \bibinfo {author} {\bibfnamefont {Mikhail~D}\
  \bibnamefont {Lukin}},\ }\bibfield  {title} {\enquote {\bibinfo {title}
  {Photon-photon interactions via rydberg blockade},}\ }\href {\doibase
  https://doi.org/10.1103/PhysRevLett.107.133602} {\bibfield  {journal}
  {\bibinfo  {journal} {Physical review letters}\ }\textbf {\bibinfo {volume}
  {107}},\ \bibinfo {pages} {133602} (\bibinfo {year} {2011})}\BibitemShut
  {NoStop}%
\bibitem [{\citenamefont {Khazali}\ \emph {et~al.}(2015)\citenamefont
  {Khazali}, \citenamefont {Heshami},\ and\ \citenamefont
  {Simon}}]{khazali2015photon}%
  \BibitemOpen
  \bibfield  {author} {\bibinfo {author} {\bibfnamefont {Mohammadsadegh}\
  \bibnamefont {Khazali}}, \bibinfo {author} {\bibfnamefont {Khabat}\
  \bibnamefont {Heshami}}, \ and\ \bibinfo {author} {\bibfnamefont {Christoph}\
  \bibnamefont {Simon}},\ }\bibfield  {title} {\enquote {\bibinfo {title}
  {Photon-photon gate via the interaction between two collective rydberg
  excitations},}\ }\href {\doibase https://doi.org/10.1038/s41567-018-0313-7}
  {\bibfield  {journal} {\bibinfo  {journal} {Physical Review A}\ }\textbf
  {\bibinfo {volume} {91}},\ \bibinfo {pages} {030301} (\bibinfo {year}
  {2015})}\BibitemShut {NoStop}%
\bibitem [{\citenamefont {Wade}\ \emph {et~al.}(2016)\citenamefont {Wade},
  \citenamefont {Mattioli},\ and\ \citenamefont {M{\o}lmer}}]{wade2016single}%
  \BibitemOpen
  \bibfield  {author} {\bibinfo {author} {\bibfnamefont {Andrew~CJ}\
  \bibnamefont {Wade}}, \bibinfo {author} {\bibfnamefont {Marco}\ \bibnamefont
  {Mattioli}}, \ and\ \bibinfo {author} {\bibfnamefont {Klaus}\ \bibnamefont
  {M{\o}lmer}},\ }\bibfield  {title} {\enquote {\bibinfo {title} {Single-atom
  single-photon coupling facilitated by atomic-ensemble dark-state
  mechanisms},}\ }\href {\doibase 10.1103/PhysRevA.94.053830} {\bibfield
  {journal} {\bibinfo  {journal} {Physical Review A}\ }\textbf {\bibinfo
  {volume} {94}},\ \bibinfo {pages} {053830} (\bibinfo {year}
  {2016})}\BibitemShut {NoStop}%
\bibitem [{\citenamefont {Sun}\ and\ \citenamefont
  {Chen}(2018)}]{sun2018analysis}%
  \BibitemOpen
  \bibfield  {author} {\bibinfo {author} {\bibfnamefont {Yuan}\ \bibnamefont
  {Sun}}\ and\ \bibinfo {author} {\bibfnamefont {Ping-Xing}\ \bibnamefont
  {Chen}},\ }\bibfield  {title} {\enquote {\bibinfo {title} {Analysis of
  atom--photon quantum interface with intracavity rydberg-blocked atomic
  ensemble via two-photon transition},}\ }\href {\doibase
  https://doi.org/10.1364/OPTICA.5.001492} {\bibfield  {journal} {\bibinfo
  {journal} {Optica}\ }\textbf {\bibinfo {volume} {5}},\ \bibinfo {pages}
  {1492--1501} (\bibinfo {year} {2018})}\BibitemShut {NoStop}%
\bibitem [{\citenamefont {Balewski}\ \emph {et~al.}(2014)\citenamefont
  {Balewski}, \citenamefont {Krupp}, \citenamefont {Gaj}, \citenamefont
  {Hofferberth}, \citenamefont {L{\"o}w},\ and\ \citenamefont
  {Pfau}}]{balewski2014rydberg}%
  \BibitemOpen
  \bibfield  {author} {\bibinfo {author} {\bibfnamefont {Jonathan~B}\
  \bibnamefont {Balewski}}, \bibinfo {author} {\bibfnamefont {Alexander~T}\
  \bibnamefont {Krupp}}, \bibinfo {author} {\bibfnamefont {Anita}\ \bibnamefont
  {Gaj}}, \bibinfo {author} {\bibfnamefont {Sebastian}\ \bibnamefont
  {Hofferberth}}, \bibinfo {author} {\bibfnamefont {Robert}\ \bibnamefont
  {L{\"o}w}}, \ and\ \bibinfo {author} {\bibfnamefont {Tilman}\ \bibnamefont
  {Pfau}},\ }\bibfield  {title} {\enquote {\bibinfo {title} {Rydberg dressing:
  understanding of collective many-body effects and implications for
  experiments},}\ }\href {\doibase 10.1088/1367-2630/16/6/063012} {\bibfield
  {journal} {\bibinfo  {journal} {New Journal of Physics}\ }\textbf {\bibinfo
  {volume} {16}},\ \bibinfo {pages} {063012} (\bibinfo {year}
  {2014})}\BibitemShut {NoStop}%
\bibitem [{\citenamefont {{\v{S}}ibali{\'c}}\ \emph {et~al.}(2017)\citenamefont
  {{\v{S}}ibali{\'c}}, \citenamefont {Pritchard}, \citenamefont {Adams},\ and\
  \citenamefont {Weatherill}}]{vsibalic2017arc}%
  \BibitemOpen
  \bibfield  {author} {\bibinfo {author} {\bibfnamefont {Nikola}\ \bibnamefont
  {{\v{S}}ibali{\'c}}}, \bibinfo {author} {\bibfnamefont {Jonathan~D}\
  \bibnamefont {Pritchard}}, \bibinfo {author} {\bibfnamefont {Charles~S}\
  \bibnamefont {Adams}}, \ and\ \bibinfo {author} {\bibfnamefont {Kevin~J}\
  \bibnamefont {Weatherill}},\ }\bibfield  {title} {\enquote {\bibinfo {title}
  {Arc: An open-source library for calculating properties of alkali rydberg
  atoms},}\ }\href {\doibase https://doi.org/10.1016/j.cpc.2017.06.015}
  {\bibfield  {journal} {\bibinfo  {journal} {Computer Physics Communications}\
  }\textbf {\bibinfo {volume} {220}},\ \bibinfo {pages} {319--331} (\bibinfo
  {year} {2017})}\BibitemShut {NoStop}%
\bibitem [{\citenamefont {Levine}\ \emph {et~al.}(2018)\citenamefont {Levine},
  \citenamefont {Keesling}, \citenamefont {Omran}, \citenamefont {Bernien},
  \citenamefont {Schwartz}, \citenamefont {Zibrov}, \citenamefont {Endres},
  \citenamefont {Greiner}, \citenamefont {Vuleti{\'c}},\ and\ \citenamefont
  {Lukin}}]{levine2018high}%
  \BibitemOpen
  \bibfield  {author} {\bibinfo {author} {\bibfnamefont {Harry}\ \bibnamefont
  {Levine}}, \bibinfo {author} {\bibfnamefont {Alexander}\ \bibnamefont
  {Keesling}}, \bibinfo {author} {\bibfnamefont {Ahmed}\ \bibnamefont {Omran}},
  \bibinfo {author} {\bibfnamefont {Hannes}\ \bibnamefont {Bernien}}, \bibinfo
  {author} {\bibfnamefont {Sylvain}\ \bibnamefont {Schwartz}}, \bibinfo
  {author} {\bibfnamefont {Alexander~S}\ \bibnamefont {Zibrov}}, \bibinfo
  {author} {\bibfnamefont {Manuel}\ \bibnamefont {Endres}}, \bibinfo {author}
  {\bibfnamefont {Markus}\ \bibnamefont {Greiner}}, \bibinfo {author}
  {\bibfnamefont {Vladan}\ \bibnamefont {Vuleti{\'c}}}, \ and\ \bibinfo
  {author} {\bibfnamefont {Mikhail~D}\ \bibnamefont {Lukin}},\ }\bibfield
  {title} {\enquote {\bibinfo {title} {High-fidelity control and entanglement
  of rydberg-atom qubits},}\ }\href {\doibase
  https://doi.org/10.1103/PhysRevLett.121.123603} {\bibfield  {journal}
  {\bibinfo  {journal} {Physical review letters}\ }\textbf {\bibinfo {volume}
  {121}},\ \bibinfo {pages} {123603} (\bibinfo {year} {2018})}\BibitemShut
  {NoStop}%
\bibitem [{\citenamefont {Du}\ \emph {et~al.}(2008)\citenamefont {Du},
  \citenamefont {Wen},\ and\ \citenamefont {Rubin}}]{du2008narrowband}%
  \BibitemOpen
  \bibfield  {author} {\bibinfo {author} {\bibfnamefont {Shengwang}\
  \bibnamefont {Du}}, \bibinfo {author} {\bibfnamefont {Jianming}\ \bibnamefont
  {Wen}}, \ and\ \bibinfo {author} {\bibfnamefont {Morton~H}\ \bibnamefont
  {Rubin}},\ }\bibfield  {title} {\enquote {\bibinfo {title} {Narrowband
  biphoton generation near atomic resonance},}\ }\href {\doibase
  10.1364/JOSAB.25.000C98} {\bibfield  {journal} {\bibinfo  {journal} {JOSA B}\
  }\textbf {\bibinfo {volume} {25}},\ \bibinfo {pages} {C98--C108} (\bibinfo
  {year} {2008})}\BibitemShut {NoStop}%
\bibitem [{\citenamefont {Liao}\ \emph {et~al.}(2014)\citenamefont {Liao},
  \citenamefont {Yan}, \citenamefont {He}, \citenamefont {Du}, \citenamefont
  {Zhang},\ and\ \citenamefont {Zhu}}]{liao2014subnatural}%
  \BibitemOpen
  \bibfield  {author} {\bibinfo {author} {\bibfnamefont {Kaiyu}\ \bibnamefont
  {Liao}}, \bibinfo {author} {\bibfnamefont {Hui}\ \bibnamefont {Yan}},
  \bibinfo {author} {\bibfnamefont {Junyu}\ \bibnamefont {He}}, \bibinfo
  {author} {\bibfnamefont {Shengwang}\ \bibnamefont {Du}}, \bibinfo {author}
  {\bibfnamefont {Zhi-Ming}\ \bibnamefont {Zhang}}, \ and\ \bibinfo {author}
  {\bibfnamefont {Shi-Liang}\ \bibnamefont {Zhu}},\ }\bibfield  {title}
  {\enquote {\bibinfo {title} {Subnatural-linewidth polarization-entangled
  photon pairs with controllable temporal length},}\ }\href {\doibase
  10.1103/PhysRevLett.112.243602} {\bibfield  {journal} {\bibinfo  {journal}
  {Phys. Rev. Lett.}\ }\textbf {\bibinfo {volume} {112}},\ \bibinfo {pages}
  {243602} (\bibinfo {year} {2014})}\BibitemShut {NoStop}%
\bibitem [{\citenamefont {Ding}\ \emph {et~al.}(2016)\citenamefont {Ding},
  \citenamefont {Wang}, \citenamefont {Zhang}, \citenamefont {Shi},
  \citenamefont {Dong}, \citenamefont {Yu}, \citenamefont {Zhou}, \citenamefont
  {Shi},\ and\ \citenamefont {Guo}}]{ding2016entanglement}%
  \BibitemOpen
  \bibfield  {author} {\bibinfo {author} {\bibfnamefont {Dong-Sheng}\
  \bibnamefont {Ding}}, \bibinfo {author} {\bibfnamefont {Kai}\ \bibnamefont
  {Wang}}, \bibinfo {author} {\bibfnamefont {Wei}\ \bibnamefont {Zhang}},
  \bibinfo {author} {\bibfnamefont {Shuai}\ \bibnamefont {Shi}}, \bibinfo
  {author} {\bibfnamefont {Ming-Xin}\ \bibnamefont {Dong}}, \bibinfo {author}
  {\bibfnamefont {Yi-Chen}\ \bibnamefont {Yu}}, \bibinfo {author}
  {\bibfnamefont {Zhi-Yuan}\ \bibnamefont {Zhou}}, \bibinfo {author}
  {\bibfnamefont {Bao-Sen}\ \bibnamefont {Shi}}, \ and\ \bibinfo {author}
  {\bibfnamefont {Guang-Can}\ \bibnamefont {Guo}},\ }\bibfield  {title}
  {\enquote {\bibinfo {title} {Entanglement between low-and high-lying atomic
  spin waves},}\ }\href {\doibase 10.1103/PhysRevA.94.052326} {\bibfield
  {journal} {\bibinfo  {journal} {Phys. Rev. A}\ }\textbf {\bibinfo {volume}
  {94}},\ \bibinfo {pages} {052326} (\bibinfo {year} {2016})}\BibitemShut
  {NoStop}%
\bibitem [{\citenamefont {Zhang}\ \emph {et~al.}(2016)\citenamefont {Zhang},
  \citenamefont {Ding}, \citenamefont {Dong}, \citenamefont {Shi},
  \citenamefont {Wang}, \citenamefont {Liu}, \citenamefont {Li}, \citenamefont
  {Zhou}, \citenamefont {Shi},\ and\ \citenamefont
  {Guo}}]{zhang2016experimental}%
  \BibitemOpen
  \bibfield  {author} {\bibinfo {author} {\bibfnamefont {Wei}\ \bibnamefont
  {Zhang}}, \bibinfo {author} {\bibfnamefont {Dong-Sheng}\ \bibnamefont
  {Ding}}, \bibinfo {author} {\bibfnamefont {Ming-Xin}\ \bibnamefont {Dong}},
  \bibinfo {author} {\bibfnamefont {Shuai}\ \bibnamefont {Shi}}, \bibinfo
  {author} {\bibfnamefont {Kai}\ \bibnamefont {Wang}}, \bibinfo {author}
  {\bibfnamefont {Shi-Long}\ \bibnamefont {Liu}}, \bibinfo {author}
  {\bibfnamefont {Yan}\ \bibnamefont {Li}}, \bibinfo {author} {\bibfnamefont
  {Zhi-Yuan}\ \bibnamefont {Zhou}}, \bibinfo {author} {\bibfnamefont {Bao-Sen}\
  \bibnamefont {Shi}}, \ and\ \bibinfo {author} {\bibfnamefont {Guang-Can}\
  \bibnamefont {Guo}},\ }\bibfield  {title} {\enquote {\bibinfo {title}
  {Experimental realization of entanglement in multiple degrees of freedom
  between two quantum memories},}\ }\href {\doibase 10.1038/ncomms13514}
  {\bibfield  {journal} {\bibinfo  {journal} {Nat. Commun.}\ }\textbf {\bibinfo
  {volume} {7}},\ \bibinfo {pages} {13514} (\bibinfo {year}
  {2016})}\BibitemShut {NoStop}%
\bibitem [{\citenamefont {Yu}\ \emph {et~al.}(2018)\citenamefont {Yu},
  \citenamefont {Ding}, \citenamefont {Dong}, \citenamefont {Shi},
  \citenamefont {Zhang},\ and\ \citenamefont {Shi}}]{yu2018self}%
  \BibitemOpen
  \bibfield  {author} {\bibinfo {author} {\bibfnamefont {Yi-Chen}\ \bibnamefont
  {Yu}}, \bibinfo {author} {\bibfnamefont {Dong-Sheng}\ \bibnamefont {Ding}},
  \bibinfo {author} {\bibfnamefont {Ming-Xin}\ \bibnamefont {Dong}}, \bibinfo
  {author} {\bibfnamefont {Shuai}\ \bibnamefont {Shi}}, \bibinfo {author}
  {\bibfnamefont {Wei}\ \bibnamefont {Zhang}}, \ and\ \bibinfo {author}
  {\bibfnamefont {Bao-Sen}\ \bibnamefont {Shi}},\ }\bibfield  {title} {\enquote
  {\bibinfo {title} {Self-stabilized narrow-bandwidth and high-fidelity
  entangled photons generated from cold atoms},}\ }\href {\doibase
  10.1103/PhysRevA.97.043809} {\bibfield  {journal} {\bibinfo  {journal} {Phys.
  Rev. A}\ }\textbf {\bibinfo {volume} {97}},\ \bibinfo {pages} {043809}
  (\bibinfo {year} {2018})}\BibitemShut {NoStop}%
\bibitem [{\citenamefont {Pritchard}\ \emph
  {et~al.}(2010{\natexlab{b}})\citenamefont {Pritchard}, \citenamefont
  {Maxwell}, \citenamefont {Gauguet}, \citenamefont {Weatherill}, \citenamefont
  {Jones},\ and\ \citenamefont {Adams}}]{Pritchard2010}%
  \BibitemOpen
  \bibfield  {author} {\bibinfo {author} {\bibfnamefont {JD}~\bibnamefont
  {Pritchard}}, \bibinfo {author} {\bibfnamefont {D}~\bibnamefont {Maxwell}},
  \bibinfo {author} {\bibfnamefont {Alexandre}\ \bibnamefont {Gauguet}},
  \bibinfo {author} {\bibfnamefont {KJ}~\bibnamefont {Weatherill}}, \bibinfo
  {author} {\bibfnamefont {MPA}\ \bibnamefont {Jones}}, \ and\ \bibinfo
  {author} {\bibfnamefont {CS}~\bibnamefont {Adams}},\ }\bibfield  {title}
  {\enquote {\bibinfo {title} {Cooperative atom-light interaction in a
  blockaded rydberg ensemble},}\ }\href {\doibase
  10.1103/PhysRevLett.105.193603} {\bibfield  {journal} {\bibinfo  {journal}
  {Phys. Rev. Lett.}\ }\textbf {\bibinfo {volume} {105}},\ \bibinfo {pages}
  {193603} (\bibinfo {year} {2010}{\natexlab{b}})}\BibitemShut {NoStop}%
\bibitem [{\citenamefont {Zhang}\ \emph {et~al.}(2011)\citenamefont {Zhang},
  \citenamefont {Chen}, \citenamefont {Liu}, \citenamefont {Loy}, \citenamefont
  {Wong},\ and\ \citenamefont {Du}}]{zhang2011optical}%
  \BibitemOpen
  \bibfield  {author} {\bibinfo {author} {\bibfnamefont {Shanchao}\
  \bibnamefont {Zhang}}, \bibinfo {author} {\bibfnamefont {JF}~\bibnamefont
  {Chen}}, \bibinfo {author} {\bibfnamefont {Chang}\ \bibnamefont {Liu}},
  \bibinfo {author} {\bibfnamefont {MMT}\ \bibnamefont {Loy}}, \bibinfo
  {author} {\bibfnamefont {George~KL}\ \bibnamefont {Wong}}, \ and\ \bibinfo
  {author} {\bibfnamefont {Shengwang}\ \bibnamefont {Du}},\ }\bibfield  {title}
  {\enquote {\bibinfo {title} {Optical precursor of a single photon},}\ }\href
  {\doibase 10.1103/PhysRevLett.106.243602} {\bibfield  {journal} {\bibinfo
  {journal} {Phys. Rev. Lett.}\ }\textbf {\bibinfo {volume} {106}},\ \bibinfo
  {pages} {243602} (\bibinfo {year} {2011})}\BibitemShut {NoStop}%
\bibitem [{\citenamefont {Ding}\ \emph {et~al.}(2015)\citenamefont {Ding},
  \citenamefont {Jiang}, \citenamefont {Zhang}, \citenamefont {Zhou},
  \citenamefont {Shi},\ and\ \citenamefont {Guo}}]{ding2015optical}%
  \BibitemOpen
  \bibfield  {author} {\bibinfo {author} {\bibfnamefont {Dong-Sheng}\
  \bibnamefont {Ding}}, \bibinfo {author} {\bibfnamefont {Yun~Kun}\
  \bibnamefont {Jiang}}, \bibinfo {author} {\bibfnamefont {Wei}\ \bibnamefont
  {Zhang}}, \bibinfo {author} {\bibfnamefont {Zhi-Yuan}\ \bibnamefont {Zhou}},
  \bibinfo {author} {\bibfnamefont {Bao-Sen}\ \bibnamefont {Shi}}, \ and\
  \bibinfo {author} {\bibfnamefont {Guang-Can}\ \bibnamefont {Guo}},\
  }\bibfield  {title} {\enquote {\bibinfo {title} {Optical precursor with
  four-wave mixing and storage based on a cold-atom ensemble},}\ }\href
  {https://journals.aps.org/prl/abstract/10.1103/PhysRevLett.114.093601}
  {\bibfield  {journal} {\bibinfo  {journal} {Physical review letters}\
  }\textbf {\bibinfo {volume} {114}},\ \bibinfo {pages} {093601} (\bibinfo
  {year} {2015})}\BibitemShut {NoStop}%
\bibitem [{\citenamefont {James}\ \emph {et~al.}()\citenamefont {James},
  \citenamefont {Kwiat}, \citenamefont {Munro},\ and\ \citenamefont
  {White}}]{james64measurement}%
  \BibitemOpen
  \bibfield  {author} {\bibinfo {author} {\bibfnamefont {Daniel~FV}\
  \bibnamefont {James}}, \bibinfo {author} {\bibfnamefont {Paul~G}\
  \bibnamefont {Kwiat}}, \bibinfo {author} {\bibfnamefont {William~J}\
  \bibnamefont {Munro}}, \ and\ \bibinfo {author} {\bibfnamefont {Andrew~G}\
  \bibnamefont {White}},\ }\bibfield  {title} {\enquote {\bibinfo {title}
  {Measurement of qubits},}\ }\href {\doibase
  https://doi.org/10.1103/PhysRevA.64.052312} {\bibfield  {journal} {\bibinfo
  {journal} {Phys Rev A}\ }\textbf {\bibinfo {volume} {64}},\ \bibinfo {pages}
  {052312}}\BibitemShut {NoStop}%
\bibitem [{\citenamefont {Wang}\ \emph {et~al.}(2017)\citenamefont {Wang},
  \citenamefont {Zhang}, \citenamefont {Zhou}, \citenamefont {Dong},
  \citenamefont {Shi}, \citenamefont {Liu}, \citenamefont {Ding},\ and\
  \citenamefont {Shi}}]{wang2017optical}%
  \BibitemOpen
  \bibfield  {author} {\bibinfo {author} {\bibfnamefont {Kai}\ \bibnamefont
  {Wang}}, \bibinfo {author} {\bibfnamefont {Wei}\ \bibnamefont {Zhang}},
  \bibinfo {author} {\bibfnamefont {Zhiyuan}\ \bibnamefont {Zhou}}, \bibinfo
  {author} {\bibfnamefont {Mingxing}\ \bibnamefont {Dong}}, \bibinfo {author}
  {\bibfnamefont {Shuai}\ \bibnamefont {Shi}}, \bibinfo {author} {\bibfnamefont
  {Shilong}\ \bibnamefont {Liu}}, \bibinfo {author} {\bibfnamefont {Dongsheng}\
  \bibnamefont {Ding}}, \ and\ \bibinfo {author} {\bibfnamefont {Baosen}\
  \bibnamefont {Shi}},\ }\bibfield  {title} {\enquote {\bibinfo {title}
  {Optical storage of orbital angular momentum via rydberg electromagnetically
  induced transparency},}\ }\href
  {https://www.osapublishing.org/col/abstract.cfm?uri=col-15-6-060201}
  {\bibfield  {journal} {\bibinfo  {journal} {Chinese Optics Letters}\ }\textbf
  {\bibinfo {volume} {15}},\ \bibinfo {pages} {060201} (\bibinfo {year}
  {2017})}\BibitemShut {NoStop}%
\end{thebibliography}
\end{document}


\title{\textcolor{black}{Supporting information of Experimental demonstration
of switching entangled photons based on the Rydberg blockade effect}}

\author{Yi-Chen Yu}

\affiliation{Key Laboratory of Quantum Information, University of Science and
Technology of China, Hefei, Anhui 230026, China.}

\affiliation{Synergetic Innovation Center of Quantum Information and Quantum Physics,
University of Science and Technology of China, Hefei, Anhui 230026,
China.}

\author{Ming-Xin Dong}

\affiliation{Key Laboratory of Quantum Information, University of Science and
Technology of China, Hefei, Anhui 230026, China.}

\affiliation{Synergetic Innovation Center of Quantum Information and Quantum Physics,
University of Science and Technology of China, Hefei, Anhui 230026,
China.}

\author{Ying-Hao Ye}

\affiliation{Key Laboratory of Quantum Information, University of Science and
Technology of China, Hefei, Anhui 230026, China.}

\affiliation{Synergetic Innovation Center of Quantum Information and Quantum Physics,
University of Science and Technology of China, Hefei, Anhui 230026,
China.}

\author{Guang-Can Guo}

\affiliation{Key Laboratory of Quantum Information, University of Science and
Technology of China, Hefei, Anhui 230026, China.}

\affiliation{Synergetic Innovation Center of Quantum Information and Quantum Physics,
University of Science and Technology of China, Hefei, Anhui 230026,
China.}

\author{Dong-Sheng Ding}
\email{dds@ustc.edu.cn}

\affiliation{Key Laboratory of Quantum Information, University of Science and
Technology of China, Hefei, Anhui 230026, China.}

\affiliation{Synergetic Innovation Center of Quantum Information and Quantum Physics,
University of Science and Technology of China, Hefei, Anhui 230026,
China.}

\author{Bao-Sen Shi}
\email{drshi@ustc.edu.cn}

\affiliation{Key Laboratory of Quantum Information, University of Science and
Technology of China, Hefei, Anhui 230026, China.}

\affiliation{Synergetic Innovation Center of Quantum Information and Quantum Physics,
University of Science and Technology of China, Hefei, Anhui 230026,
China.}

\date{\today}

\maketitle
\textbf{Experimental time sequence. }The repetition rate of our experiment
is $200$~Hz, and the MOT trapping time is 4.71~ms. The experimental
time window is 25~$\mu$s. The magnetic field is switched off during
the switch process. The fields of pumps 1 and 2 are controlled by
two AOMs, and therefore the frequencies of signals 1 and 2 photons
can be tuned. The optical depth in MOT 1 is about 20. Each of the
two lenses with a focal length of 300~mm, is used to couple the signal
fields into the atomic ensemble in MOT 1. The short-focus lens (the
blue lens in figure 1) with a focal length of 30~mm is used to couple
the signal-2 field into the atomic ensemble in MOT 2. The fields of
P1 and P2 are collinear, and the signal fields S1 and S2 are collinear.
The phase matching condition $k_{p1}-k_{s1}=k_{p2}-k_{s2}$ is satisfied
in the spontaneous four-wave mixing process. The two signal photons
are collected into their respective single-mode fibers and are detected
by two single-photon detectors (avalanche diode, PerkinElmer SPCM-AQR-16-FC,
60\% efficiency, maximum dark count rate of 25/s). The two detectors
are gated by an arbitrary function generator. The gated signals from
the two detectors are then sent to a time-correlated single-photon
counting system (TimeHarp 260) to measure their time-correlated function.

\textbf{Generating entangled photons}\textcolor{black}{. We generate
entangled photon pairs in MOT 1 which is presented in the upper half
of fig.1 (c). We generate the entangled photons via SFWM process by
using the counter-propagating pump fields 1 and 2. The SFWM process
is based on double-\textgreek{L} atomic configuration with energy
levels of ground state} $\left|1\right\rangle $\textcolor{black}{,
metastable state} $\left|3\right\rangle $\textcolor{black}{, excited
states} $\left|2\right\rangle $\textcolor{black}{{} and} $\left|4\right\rangle $\textcolor{black}{,
which correspond to 5S}\textsubscript{\textcolor{black}{1/2}}\textcolor{black}{(F=2),
5S}\textsubscript{\textcolor{black}{1/2}}\textcolor{black}{(F=3),
5P}\textsubscript{\textcolor{black}{1/2}}\textcolor{black}{(F'=3)
and 5P}\textsubscript{\textcolor{black}{3/2}}\textcolor{black}{(F'=3).
Pump 1 at 795 nm with horizonal polarization is from an external-cavity
diode laser (DL100, Toptica). Pump 2 at 780 nm with vertical polarization
into the atomic medium is from anoter external-cavity diode laser
(DL100, Toptica). These two pump lasers with different polarizations
are propagating in a strictly collinear geometry, with an angle of
2${^\circ}$ away from the direction of collected photons \cite{liao2014subnatural}.
Signal 1 and signal 2 photons are non-classically correlated \cite{ding2012generation}.
Afterwards, we form a phase self-stabilized multiplexing structure
where the relative phase between different signal paths can be eliminated
because of the symmetrical structure, by using the beam displacer
1, beam displacer 2 and two half-wave plates. Then, signal 1 and signal
2 photons are not only non-classical correleted but also polarization
entangled. The form of the entanglement is $\left|\psi\right\rangle =(\left|H_{s1}\right\rangle \left|V_{s2}\right\rangle +e^{i\theta}\left|V_{s1}\right\rangle \left|H_{s2}\right\rangle )/\sqrt{2}$.
In Ref \cite{Yu2018}, we have already checked the self-stabilization
of our structure and the high fidelity of polarization entanglement.
We can produce all four Bell states by adjusting the two BDs.}

\textbf{Theoretical analysis.} We use the Lindblad master equation:
$d\rho/dt=-i[H,\rho]/\hbar+L/\hbar$ in the absence of Rydberg-mediated
interactions to characterize the interaction of the EIT light fields
with an ensemble of three level atoms, where $\rho$ is the the atomic
ensemble's density matrix and $H$ is the atom-light interaction Hamiltonian
summed over all the single-atom Hamiltonians under rotating wave approximation
with $H=\sum_{k}H[\rho^{(k)}]$.
\begin{center}
$H[\rho^{(k)}]=-\frac{1}{2}\hbar\left(\begin{array}{ccc}
0 & \Omega_{p} & 0\\
\Omega_{p} & -2\Delta_{p} & \text{\textgreek{W}}_{c}\\
0 & \text{\ensuremath{\text{\textgreek{W}}_{c}}} & -2(\text{\ensuremath{\Delta}}_{p}+\text{\ensuremath{\Delta}}_{c})
\end{array}\right)$
\par\end{center}

The Lindblad superoperator $L=\sum_{k}L[\rho^{(k)}]$ is comprised
of the single-atom superoperators. The Lindblad master equation includes
both spontaneous emissions and dephasing. $\Omega_{p}$ and $\Omega_{c}$
are the Rabi frequency of the probe and coupling light. $\Delta_{p}$and
$\Delta_{c}$ are the detuning of probe and coupling light, respectively.
Considering the dephasing of Rydberg state, we introduce $\gamma_{d}$
to characterize the decoherence. Since the dephasing of Rydberg state
doesn't include population transfer, it ought to be included only
as a decay of the coherence, that is in the off-diagonal terms of
the Lindblad operator. The single-atom Lindblad superoperator is :

$2L[\rho^{(k)}]/\hbar=$

\begin{figure}[h]
\includegraphics[width=1\columnwidth]{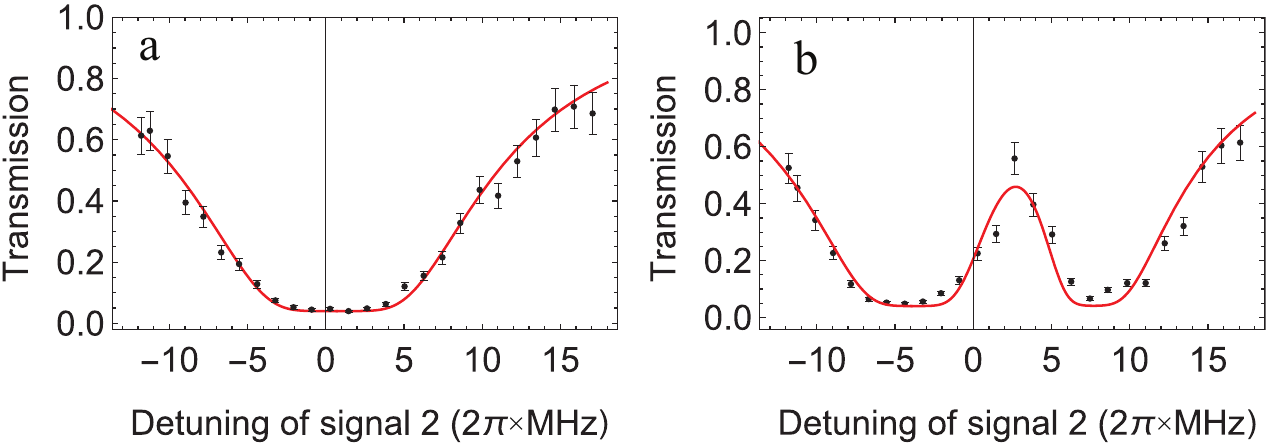}

\caption{Transmission of the signal-2 photon under different conditions. (a)
Absorption spectra of the signal-2 photon with atoms present in MOT
2. The data are fitted by function $e^{-2\mathrm{Im}[w_{s2}/c(1+\chi/2)]L}+a_{0}$
(solid lines) with parameter values of $\Omega_{c}=0$, $\gamma_{eg}=2\pi\times3$~MHz,
$\gamma_{rg}=2\pi\times1$~MHz, $OD=20$, and $a_{0}=0.04$. (b)
Rydberg-EIT effect of the signal-2 photon with coupling field present.
The data are fitted by the same function above with parameter values
of $\Omega_{c}=2\pi\times11$~MHz, $\gamma_{eg}=2\pi\times3$~MHz,
$\gamma_{rg}=2\pi\times0.1$~MHz, $OD=20$, $a_{0}=0.04$ and detuning
shift $\delta\Delta=-2\pi\times3.2$ MHz.}

\label{EIT}
\end{figure}

\textbf{
\[
\left(\begin{array}{ccc}
2\mathbf{\Gamma}_{e}\rho_{ee}^{(k)} & -\text{\ensuremath{\mathbf{\Gamma}}}_{e}\rho_{ge}^{(k)} & -\text{\ensuremath{\mathbf{\Gamma}}}_{r}\rho_{gr}^{(k)}\\
-\mathbf{\Gamma}_{e}\rho_{eg}^{(k)} & -2\text{\ensuremath{\mathbf{\Gamma}}}_{e}\rho_{ee}^{(k)}+2\text{\ensuremath{\mathbf{\Gamma}}}_{r}\rho_{rr}^{(k)} & -(\text{\ensuremath{\mathbf{\Gamma}}}_{e}+\text{\ensuremath{\mathbf{\Gamma}}}_{r})\rho_{er}^{(k)}\\
-\text{\ensuremath{\mathbf{\Gamma}}}_{r}\rho_{rg}^{(k)} & -(\text{\ensuremath{\mathbf{\Gamma}}}_{e}+\text{\ensuremath{\mathbf{\Gamma}}}_{r})\rho_{re}^{(k)} & -2\text{\ensuremath{\mathbf{\Gamma}}}_{r}\rho_{rr}^{(k)}
\end{array}\right)
\]
+
\[
\left(\begin{array}{ccc}
0 & -\text{\ensuremath{\gamma}}_{de}\rho_{ge}^{(k)} & 0\\
-\text{\ensuremath{\gamma}}_{de}\rho_{eg}^{(k)} & 0 & -\text{\ensuremath{\gamma}}_{de}\rho_{er}^{(k)}\\
0 & -\text{\ensuremath{\gamma}}_{de}\rho_{re}^{(k)} & 0
\end{array}\right)
\]
+
\[
\left(\begin{array}{ccc}
0 & 0 & -\text{\ensuremath{\gamma}}_{dr}\rho_{gr}^{(k)}\\
0 & 0 & -\text{\ensuremath{\gamma}}_{dr}\rho_{er}^{(k)}\\
-\text{\ensuremath{\gamma}}_{dr}\rho_{rg}^{(k)} & -\text{\ensuremath{\gamma}}_{dr}\rho_{re}^{(k)} & 0
\end{array}\right)
\]
}

where $\text{\ensuremath{\mathbf{\Gamma}}}_{e}$ and $\text{\ensuremath{\mathbf{\Gamma}}}_{r}$
are the natural population decay rates of excited state $\left|e\right\rangle $
and Rydberg state $\left|r\right\rangle $, respectively. $\text{\ensuremath{\gamma}}_{de}$
and $\text{\ensuremath{\gamma}}_{dr}$ are the decay of the coherence
of excited state $\left|e\right\rangle $ and Rydberg state $\left|r\right\rangle $
caused by collisions and stray fields. We solve the Lindblad master
equation under the condition of steady state corresponding to $t\rightarrow\infty,d\rho/dt=0$
to get $\rho_{eg}$. The complex susceptibility of the EIT medium
is $\chi=(N|\mu_{ge}|^{2}/\epsilon_{0}\hbar)\rho_{eg}$ \cite{Harris1990},
with $N$ being the atom number density in the medium, and $\mu_{ge}$
being the electric dipole moment of the state $\left|g\right\rangle \rightarrow\left|e\right\rangle $.
Under the plane-wave approximation, the transmission of the signal-2
photon through the EIT medium can be obtained from the susceptibility
via $e^{-2\mathrm{Im}[w_{s2}/c(1+\chi/2)]L}$, where $L$ is the length
of the atomic medium, $w_{s2}$ the frequency of the signal-2 photon,
$c$ the speed of light in a vacuum. We calculate the complex linear
susceptibility \cite{zhang2012dark} :

$\chi=$
\begin{equation}
\frac{\alpha_{0}}{k_{0}}\frac{4(\bigtriangleup w_{s2}+\delta\Delta+i\gamma_{rg})\gamma_{eg}}{\Omega_{c}^{2}-4(\bigtriangleup w_{s2}+\delta\Delta+i\gamma_{rg})(\bigtriangleup w_{s2}+\delta\Delta+i\gamma_{eg})}\label{eq:EIT}
\end{equation}
where $\alpha_{0}=OD/L$ is the absorption coefficient when the coupling
field is not present, $OD$ is the optical depth of atomic ensemble.
$k_{0}$ is the wave vector. $\bigtriangleup w_{s2}$ is the detuning
of the signal-2 photon. $\delta\Delta$ is the frequency shift used
for fitting EIT spectrum. $\gamma_{eg}=\text{\ensuremath{\mathbf{\Gamma}}}_{e}+\text{\ensuremath{\gamma}}_{de}$
is the decay rate of atomic transition $\left|e\right\rangle \rightarrow\left|g\right\rangle $
corresponding to the natural linewidths of $\left|e\right\rangle $.
$\gamma_{rg}=\text{\ensuremath{\text{\ensuremath{\mathbf{\Gamma}}}_{r}}+\ensuremath{\gamma}}_{dr}$
is the decay rate of atomic transition $\left|nD\right\rangle \rightarrow\left|g\right\rangle $
\cite{Raitzsch2008Investigation}. $\Omega_{c}$ represents the Rabi
frequency of the coupling field. We use this equation \ref{eq:EIT}
to simulate the results given in figure~\ref{EIT} (a) and (b). The
strong dipole interaction couples the nearby Rydberg atoms so that
the evolution of these atoms are fundamentally linked, thereby modifying
the individual atomic energy levels \cite{keaveney2014collective}.
As a result of the Rydberg dipole interactions, the behaviour of an
ensemble of $N$-atoms cannot simply be described by summing the response
of a single atom $N$ times. To describe the behavior for Rydberg-EIT
with a gate field, we introduce $\gamma_{d}$ to characterize the
decoherence to fit the data from our experimental result. If there
are no Rydberg dipole interactions, the transmission of the $N$-atom
system can be traced to a summation of $N$ single-atom contributions.
$\gamma_{d}$ would increase when the Rydberg atoms interact, and
the response of each atom is modified significantly. Because of this
process, the response of Rydberg atoms would exhibit non-transparency
behavior for signal photons when $\gamma_{dr}$ is large enough. Thus,
we can control the single-photon transmission behavior of Rydberg-EIT
depending on whether nearby Rydberg atoms are excited. In our experiment,
the coupling Rabi frequency is higher than the polarization decay
rate of the excited state but not much higher enough to absolutely
separate the two absorption spectra \cite{Abi2007Interference}. Furthermore,
we have checked the linear susceptibility under the condition of either
EIT or AT effect, confirming that the result can be used in all value
of $\Omega_{c}$. Our scheme of switching an entangled photon could
work well under both EIT and AT conditions.

\textbf{Frequency matching between signal 2 and atoms in MOT 2}. We
need to match the frequency windows to connect two different physical
systems. For this point, the emitted signal-2 photon from MOT 1 may
not be matched with the working window of Rydberg-EIT in MOT 2. The
detuning of the signal-2 photon is performed by changing the frequency
of the pump-2 field, which is controlled by an AOM. The pump 2 field
passes through the AOM with a frequency shift from $-2\pi\times12\sim2\pi\times17$~MHz.
There is another AOM added in the optical path to give the signal-2
photon (see figure~1 (c) in main text) a frequency shift with $+2\pi\times120$~MHz.
By this method, the frequency of the signal-2 photon can be tuned
from the atomic transition $5S_{1/2}(F=3)\rightarrow5P_{3/2}(F'=3)$
to the atomic transition $5S_{1/2}(F=3)\rightarrow5P_{3/2}(F'=4)$.
To check whether the signal-2 photon falls into the atomic transition
window $5S_{1/2}(F=3)\rightarrow5P_{3/2}(F'=4)$ in MOT 2, we measure
its absorption and EIT transmission spectra (figure~3) by changing
the frequency of the emitted signal-2 photon. To check this process,
we added a coupling field, which is resonant with the atomic transition
$5P_{3/2}(F'=3)\rightarrow50D_{5/2}$ to demonstrate the Rydberg-EIT.

\begin{figure}[h]
\includegraphics[width=1\columnwidth]{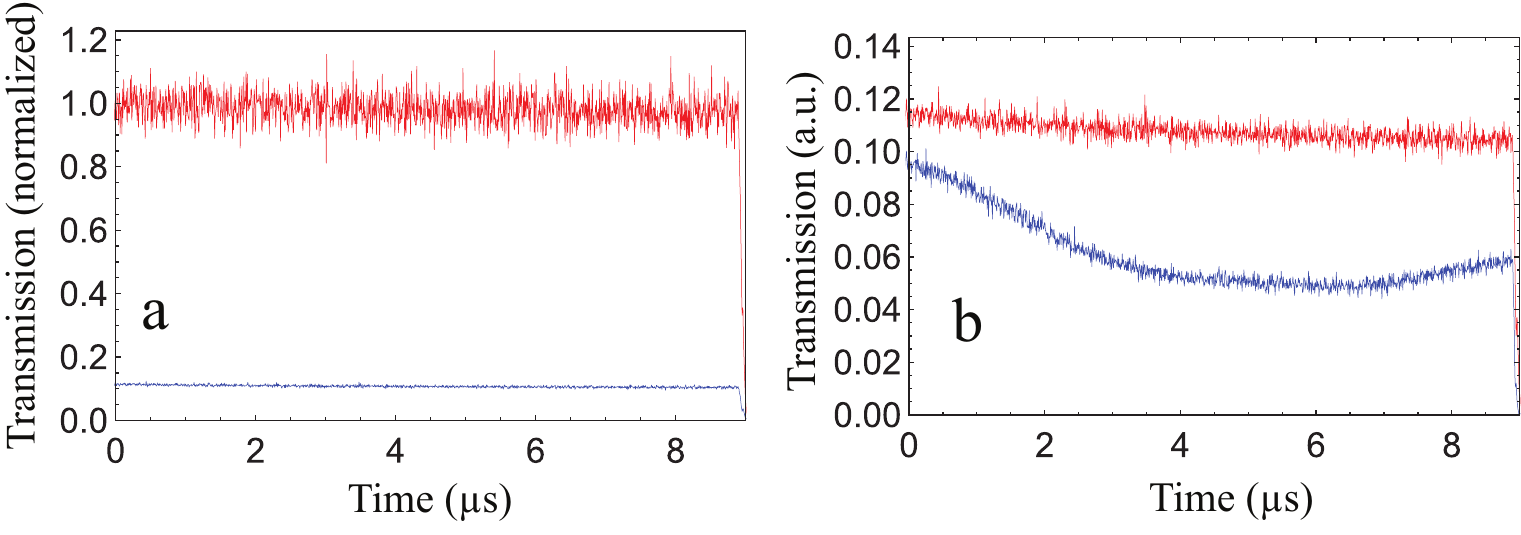}

\caption{(a) The normalized transmission of probe field on Rydberg-EIT resonance.
The red line represents 1.5 photons/$\mu s$, the blue line corresponds
to 38.7 photons/$\mu s$. (b) Time dependence of probe transmission
with different photons (not normalized, only see the profile of the
transmission line). The red line represents 38.7 photons/$\mu s$,
the blue line corresponds to 601.4 photons/$\mu s$.}

\label{pulse}
\end{figure}

\textbf{Time dependence of the transmission spectrum of Rydberg-$D$
state.} The Ref. \cite{Tresp2015} has reported the dephasing of Rydberg-$D$
state polaritons, namely, time dependence of the transmission on EIT
resonance. As stated in this reference, the Rydberg-$D$ state dephasing
effect is caused by the interaction-induced coupling to degenerate
Zeeman sublevels, preventing other photons propagating through the
cloud. We also checked the $D$-state dephasing in our experiment
to demonstrate probe field transmission under the EIT resonance condition.
We use a probe field with 1.5 photons/$\mu s$ as an input field,
the transmission is higher than using 38.7 photons/$\mu s$, (see
the results given in Fig. \ref{pulse} (a)). We want to show that
the gate field with 1.5 photon/$\mu s$ is not enough to reach a strong
quantum nonlinearity to obviously destroy the transparency. Thereby,
the transmission rate is almost 1 (actually lower than 1). But the
strong interaction between Rydberg atoms could easily tunes the transition
out of resonance on the condition of the gate field with 38.7 photons/$\mu s$.
The comparison is consistent with the result of our experiment. And
the gate field is almost 15.5 photons/$\mu s$ during our experiment.
We use the normalized transmission in the left panel to distinguish
the transmission rate under two different conditions. The Rydberg-$D$
state dephasing is not obvious for small photon numbers, but with
an obvious time dependence with much more photon numbers, given in
Fig. \ref{pulse} (b). So, the $D$-state dephasing caused by the
gate field is not obvious in our experiment, for the gate photon number
we used in our experiment is 15.5 photons/$\mu s$, which is smaller
than 38.7 photons/$\mu s$. In addition, we also compared the Rydberg-EIT
effects for $D$ state and $S$ state by tuning the coupling laser
wavelength, the only difference between them in our configuration
is the height of the transparent peak with the transparent peak of
$D$ state higher than that of $S$ state. In order to obviously observe
the blockade effect in Rydberg-EIT, we choose Rydberg-$D$ state to
demonstrate the experiment.

\begin{figure}[h]
\includegraphics[width=1\columnwidth]{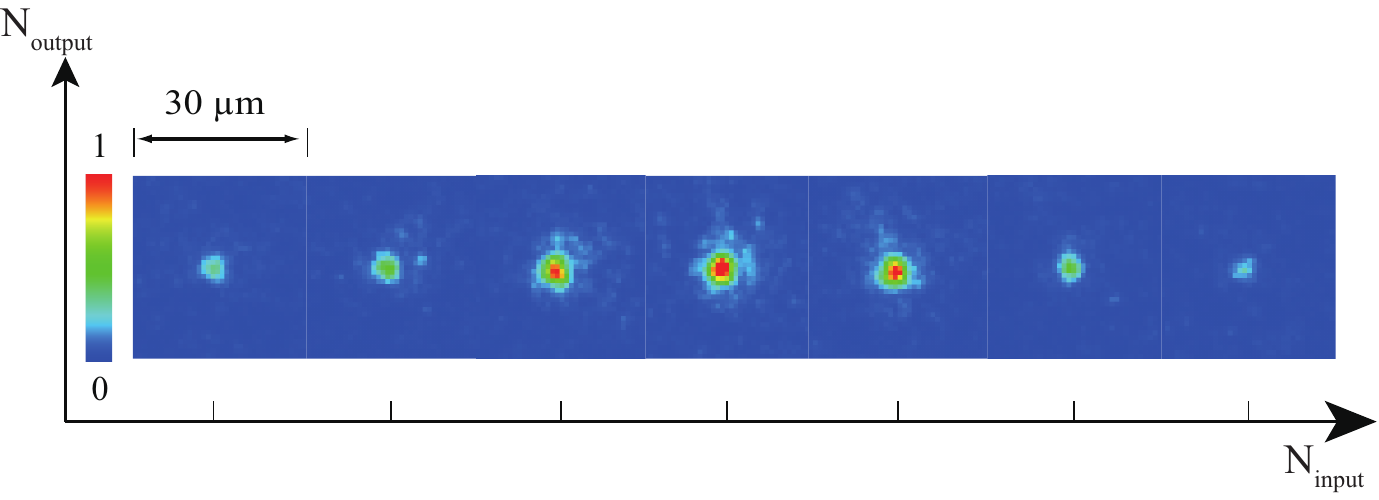}

\caption{We\textcolor{black}{{} change the input photon numbers N}\protect\textsubscript{\textcolor{black}{input}}\textcolor{black}{{}
from a week coherent light beam. Then we record the output photon
numbers N}\protect\textsubscript{\textcolor{black}{output}}\textcolor{black}{{}
after the propagation through the MOT 2 proving that N}\protect\textsubscript{\textcolor{black}{output}}\textcolor{black}{{}
nonlinearly increase as N}\protect\textsubscript{\textcolor{black}{input}}\textcolor{black}{{}
increase.}}

\label{nonlinear}
\end{figure}

\textbf{\textcolor{black}{Nonlinear response of Rydberg blockaded
ensemble.}}\textcolor{black}{{} We provide the visualized evidence of
nonlinear response of Rydberg blockaded ensemble. Before the switch
operation in our experiment, we change the input photon numbers (N}\textsubscript{\textcolor{black}{input}}\textcolor{black}{)
of signal-2 photons which are from a week coherent light beam. Then
we record the output photon numbers (N}\textsubscript{\textcolor{black}{output}}\textcolor{black}{)
after the propagation through the MOT 2 Rydberg-EIT transparent window
by a fast and high-efficient camera (PI-MAX 4, princeton instruments)
shown in Fig. \ref{nonlinear}. As we increase N}\textsubscript{\textcolor{black}{input}}\textcolor{black}{,
N}\textsubscript{\textcolor{black}{output}}\textcolor{black}{{} nolinearly
increase or even decrease because of the Rydberg-blockade effect.
Thus, there is a threshold in our Rydberg ensemble. If the input photon
numbers exceed that threshold, our ensemble exists large nolinearity
to block the input photons. In fact, the gate field is applied to
our ensemble to approach that threshold.}